\documentclass[onecolumn,usenatbib,amsmath,amssymb,amsbsy]{mn2e}

\usepackage{amsmath,amssymb,amsbsy,amsfonts}

\usepackage{natbib}
\usepackage{graphicx}

\newcommand{\be}{\begin{equation}}
\newcommand{\ee}{\end{equation}}
\newcommand{\bear}{\begin{eqnarray}}
\newcommand{\eear}{\end{eqnarray}}

\newcommand{\rx}{{\rm x}}
\newcommand{\ry}{{\rm y}}
\newcommand{\rn}{{\rm n}}
\newcommand{\re}{{\rm e}}
\newcommand{\rp}{{\rm p}}
\newcommand{\rxy}{{\rm xy}}
\newcommand{\ryx}{{\rm yx}}
\newcommand{\rpn}{{\rm pn}}
\newcommand{\rnp}{{\rm np}}
\newcommand{\rpe}{{\rm pe}}

\newcommand{\rne}{{\rm ne}}

\newcommand{\rv}{{\rm v}}

\newcommand{\en}{\varepsilon_\rn}
\newcommand{\ep}{\varepsilon_\rp}

\newcommand{\epstar}{\varepsilon_\star}
\newcommand{\mut}{\tilde{\mu}}

\newcommand{\cW}{{\cal W}}
\newcommand{\cE}{{\cal E}}
\newcommand{\cR}{{\cal R}}
\newcommand{\cC}{{\cal C}}
\newcommand{\cT}{{\cal T}}

\newcommand{\cN}{{\cal N}}

\newcommand{\cL}{{\cal L}}

\begin{document}

\title[MHD of neutron star cores]
{Magnetohydrodynamics of superfluid and superconducting neutron star cores}

\author[Glampedakis, Andersson \& Samuelsson]{Kostas Glampedakis$^{1}$,  Nils Andersson$^2$ and Lars Samuelsson$^{3,4}$ \\
\\
$^1$ Theoretical Astrophysics, Auf der Morgenstelle 10, University of Tuebingen, D-72076 Tuebingen, Germany \\
$^2$ School of Mathematics, University of Southampton, Southampton SO17 1BJ, UK \\
$^3$ Department of Physics, Ume\aa\ University, SE-901 87 Ume\aa, Sweden \\
$^4$ Nordita, Roslagstullsbacken 23, SE-106 91 Stockholm, Sweden}

\maketitle

\begin{abstract}
Mature neutron stars are cold enough to contain a number of superfluid and superconducting components.
These systems are distinguished by the presence of additional dynamical degrees of freedom associated with
superfluidity.
In order to consider models with mixtures of condensates we need to develop a multifluid
description that accounts for the presence of rotational neutron vortices and magnetic proton fluxtubes. We also need to
model the forces that impede the motion of vortices and fluxtubes, and understand how these forces
act on the condensates. This paper concerns the development of such a model for the outer core
of a neutron star, where superfluid neutrons co-exist with a type~II proton superconductor and
an electron gas. We discuss the hydrodynamics of this system, focusing on the role of the entrainment effect,
the magnetic field,
the vortex/fluxtube tension and the dissipative mutual friction forces. Our final results can be directly
applied to a number of interesting astrophysical scenarios, e.g. associated with neutron star oscillations or
the evolution of the large scale magnetic field.
\end{abstract}


\section{Introduction}
\label{sec:intro}

Neutron stars host the strongest magnetic fields known in nature. The magnetic field of a typical
pulsar has an (already) enormous intensity of $\sim 10^{12}\, \mbox{G} $, but this is nothing compared to
magnetars which can reach, or even exceed, the extraordinary intensity of $\sim 10^{15}\, \mbox{G} $ \citep{DT}.
Given this, it is not surprising that the magnetic field plays a key role in neutron star physics.
As far as neutron star dynamics is concerned, the presence of the magnetic field is likely to affect large-scale
phenomena like pulsar glitches and free precession. It may also alter
the properties of the star's various oscillation modes. In fact, if the field is sufficiently strong, it may provide the main
restoring force acting on the fluid elements, leading to Alfv\'en-type oscillation modes. Such effects may already be
observed in quasi-periodic oscillations associated with the tails of magnetar giant flares
(see \cite{stroh_review} for a review).

In its vast majority,  existing work on the dynamics of magnetised neutron stars makes use of
standard magnetohydrodynamics (MHD)  (see, for example, \cite{STbook}) which is based on a ``single-fluid'' hydrodynamical model.
Although this is a reasonable initial approach to the neutron star problem, it is of questionable validity
in more realistic scenarios. This is easy to see if we recall that below a critical temperature threshold
$ T_c \sim 5 \times 10^9\,\mbox{K} $ (promptly reached after a neutron star's birth) the liquid matter in neutron star interiors is
expected to undergo a phase transition to a quantum ``ordered'' state where the bulk of the neutrons (protons) become superfluid
(superconducting) through the formation of Cooper pairs.
As a result of this phase transition the macroscopic hydrodynamical behaviour of neutron star matter is that of a
multifluid system with more than one distinct fluid degree of freedom. In addition, the presence of a superconductor in neutron star
matter means that the properties of the macroscopic magnetic field are likely to differ from the familiar plasma physics behaviour.

The number of distinct ``fluids'' appearing below $ T_c$ depends on the composition of matter. Since
the outer core of a
canonical neutron star is primarily composed of neutron-proton-electron (npe) matter, one can
identify three fluids\footnote{This is strictly true only if we ignore finite temperature effects. Otherwise the entropy
associated with finite temperature excitations may provide an additional degree of freedom \citep{prix04,nahelium,heatpaper}.}. 
The key point is that superfluidity allows for relative flows among interpenetrating matter components.
In the inner neutron star core the situation could be further complicated
by the presence of exotic particle species like hyperons or deconfined quarks. However, even though this leads to very 
interesting problems, we will not consider the deep core in this paper.

A consistent formalism for neutron star magnetohydrodynamics must be based on a
multifluid model which accounts for superfluidity and superconductivity. The main aim of the present paper is
to formulate such a model. Naturally, this subject has been addressed in the past, in particular
in a series of papers by Mendell and
Lindblom \citep{menlin,mendell,mendell98}. Our analysis builds on their pioneering work.

We consider a neutron star as
a three fluid system (superfluid neutrons, superconducting protons and normal electrons)
at zero temperature, coupled through entrainment, mutual friction forces and the magnetic field.
Despite this common ground with \citet{menlin} we present a substantial revision of the subject.
In particular, we correct a conceptual error in the original calculation by \citet{mendell,mendell98},
regarding the energy contribution of the so-called London field (the characteristic magnetic field induced
in a rotating superconductor). This error was first pointed out by \citet{carter}, although they did not attempt to
reformulate the MHD problem. As we will show in this paper, this error has no serious repercussions
for most of the final neutron star MHD equations. The reason for this is the intrinsic weakness (several
orders of magnitude smaller than the stellar magnetic field) of the London field. Hence, most of our 
final equations are similar to those of \citet{mendell,mendell98}. A notable exception is
Amp\'ere's law, which will be shown to depart significantly from its usual form as a result of the London
field.

The main new contribution of our work is the derivation of the magnetic force  in the MHD
equations for superfluid/superconducting neutron stars. In particular, we provide a general proof that the standard
electromagnetic Lorentz force is eliminated from these equations, being replaced by a tension force due to
the proton vortices (fluxtubes). This result was also obtained by \citet{mendell,mendell98}, although it was demonstrated only
for a plane-wave model. In addition to the derivation of the superconducting magnetic force, we provide a detailed discussion 
on the various vortex-mediated  mutual friction forces that appear in the equations of motion. For each of these forces 
we provide an explicit (albeit phenomenological) expression and discuss its properties and expected strength.
This analysis builds on previous work by \citet{mendell}.

The final difference between our work and that of \citet{mendell,mendell98} concerns the choice of variables.
We do not use the
``orthodox'' approach to superfluid hydrodynamics, where the superfluid ``velocities'' in fact represent momenta
\citep{khalatnikov}.
Instead, we prefer to use the canonical formulation developed by Carter and collaborators
\citep{carter95, prix04, monster} which makes a clear distinction between kinematical velocities and momenta, avoiding 
conceptual confusion. Since the two approaches are mathematically equivalent (see, for example, \citet{prix04}), opting for one
or the other is a matter of taste. Yet, we believe that the variational formulation offers a deeper insight into the 
physics of multifluid systems. In particular, it uses conserved fluxes, see \citet{nahelium}. This approach is also advantageous since it is 
readily extended to fully general relativistic models \citep{livrev}.


\section{An overview of neutron star superconductivity}
\label{sec:supercon}

To set the stage for the discussion of the MHD of superfluid and superconducting
neutron stars, we first provide a qualitative overview of neutron star superconductivity.
Our discussion  draws heavily on the classic work of \citet{baym} and \citet{mendell},
which in many ways still represents the state-of-the-art in this area. However, we also
touch upon some modern aspects of superconductivity.

The effect of superconductivity on neutron star dynamics is two-fold. Firstly, the protons decouple from the
rest of the matter and provide a distinct macroscopic fluid degree of freedom, in a manner analogous to the
superfluid neutrons. Secondly, the stellar magnetic field exhibits a  ``non-classical'' behaviour due to the
Meissner effect, the expulsion of magnetic flux from superconducting matter. In order to formulate the MHD equations for
superconducting neutron stars it is  necessary to  understand the implications of these effects.

One may expect to find many similarities between the well established physics of terrestrial superconductors and neutron stars.
In reality, this is only partially the case. Laboratory systems exhibit a complete flux expulsion due to the Meissner effect
for any magnetic field strength below a critical threshold. This is not the case for neutron stars, where the superconducting matter
can be penetrated by any finite magnetic field. Basically, the magnetic field is metastable with the
global Meissner effect operating on a secular timescale (of the order of Myrs), due to the high ambient electric conductivity of
neutron star matter \citep{baym}.

The magnetic flux penetrates the matter by redistributing itself into filamentary structures within which proton superconductivity
is suppressed. The topology and properties of the resulting regions, occupied by normal protons, depend on the \underline{type}
of superconductivity \citep{tilley,tinker}. The familiar dichotomy between type I and type II superconductivity also applies for
neutron stars (at least when there is a single superconducting particle species).
Which type of superconductivity prevails depends on the relative size of two characteristic lengthscales: the proton coherence
length $\xi_\rp$, on which Cooper-pairing can be broken by proton quantum fluctuations, and the penetration length $\Lambda_* $,
on which the magnetic field is exponentially shielded inside superconducting matter. The type of superconductivity
is determined by the ratio
\be
\kappa_s = \frac{\xi_\rp}{\sqrt{2} \Lambda_*}
\ee
In the case of type II superconductivity $\kappa_s < 1 $ and it is energetically favourable for the magnetic field
to penetrate the superconductor by forming an array of quantised proton vortices\footnote{Rather than
using the terminology proton fluxtubes (fluxoids) and neutron vortices, we will refer to both as ``vortices'' 
associated with each fluid. This makes sense since, in the neutron star case, the neutron vortices are magnetised. 
Moreover, both sets follow from the quantization of the (canonical) momentum circulation.}, each carrying a single flux quantum
$\phi_0 = hc/2e \sim 2\times 10^{-7}$~Gcm$^2$. Locally, each proton vortex has a cylindrical geometry with a normal proton core of
radius $\xi_\rp $. The detailed structure of proton vortices in neutron stars has been discussed by \citet{mendell}, who finds
\be
\xi_\rp \approx 5 \times 10^{-12} \left ( \frac{m_\rp}{m^*_\rp} \right )
\rho_{14}^{1/3} \left ( \frac{x_\rp}{0.1} \right )^{1/3} \left ( \frac{10^9\,{\rm K}}{T_{\rm c p}} \right )\,
{\rm cm}
\label{coher}
\ee
and (see Appendix)
\be
 \Lambda_* \approx 9 \times 10^{-12} \left ( \frac{m_\rp^*}{m_\rp} \right )^{1/2} \rho^{-1/2}_{14}
\left ( \frac{0.1}{x_\rp} \right )^{1/2}  \, {\rm cm}
\label{Lambda}
\ee
where $\rho_{14} = \rho/ 10^{14}\,{\rm g}\, {\rm cm}^{-3} $ is the normalised mass density, $x_\rp $ is
the proton fraction, $m_\rp$ is the bare proton mass and $m_\rp^* $ is the effective mass acquired
by each proton as a result of the entrainment.
Finally, $T_{\rm cp}$ is the (density dependent) critical temperature for proton superconductivity.

As in a laboratory superconductor, the upper critical field $H_{\rm c2} $  is given by \citep{tinker},
\be
H_{\rm c 2} = \frac{\phi_0}{2\pi\xi_\rp^2} \approx 10^{15} \left({m_\rp^* \over m_\rp}\right)
\left({x_\rp \over 0.1} \right)^{-2/3} \rho_{14}^{-2/3}
\left({T_\mathrm{cp} \over 10^9\ \mathrm{K}} \right)^2  \ \mathrm{G}
\label{Hc2}
\ee
This corresponds to a state where the proton vortices are densely packed to the point where
their cores ``touch'', leading to a global destruction of superconductivity.
That is, we have
\be
\xi_\rp \approx d_\rp \sim \left({\phi_0 \over H_\mathrm{c2}} \right)^{1/2}
\ee
where $d_\rp$ is the proton vortex separation.

\begin{figure}
\centering
\includegraphics[width=10cm,clip]{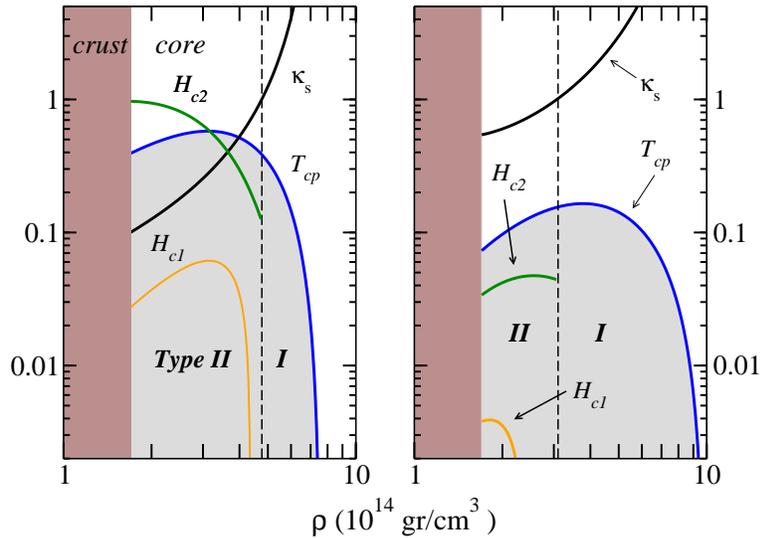}
\caption{This figure provides a superconductivity ``cross section'' of a mature neutron star core (assuming npe composition) as a function
of the density. We plot the critical temperature $ T_{\rm cp} $ for proton condensation (normalised to $10^{10}$ K),
the critical fields $ H_{\rm c2},~H_{\rm c1} $  (normalised to $10^{16} $ G) for type II superconductivity,
and the ratio $\kappa_s $ which determines the type of superconductivity. Note that the vertical axis is dimensionless.
The left (right) panel corresponds to a model of strong (weak) proton superfluidity (see text for details). 
The vertical dashed line indicates the $\rho_{\rm crit} $ density for the type II $\to$ I transition.}
\label{fig:type}
\end{figure}

In the laboratory, the vortex state is only accessible for fields in the range
$H_{\rm c1} < B < H_{\rm c2} $, where the lower critical field is given by
\be
H_\mathrm{c1} = { \phi_0 \over 4 \pi \Lambda_*^2} \log \left( { \Lambda_* \over \xi_\rp } \right)
\ee
Taking $(1/2)\log \left( { \Lambda_*/ \xi_\rp } \right)\approx 1$, cf. \citet{tinker}, we have
\be
H_\mathrm{c1} \approx { \phi_0 \over 2 \pi \Lambda_*^2} \approx  4\times 10^{14} \left({m_\rp \over m_\rp^*} \right)
\left({x_\rp \over 0.1} \right) \rho_{14} \ \mathrm{G}
\label{Hc1def}
\ee
Each proton vortex is associated with a magnetic field $\bar{B}^i_\rp $ that
points along the (local) direction of the vortex core and is distributed within a
radius $\sim \Lambda_* $. That is, we have
\be
\bar{B}_\rp \sim \frac{\phi_0}{\pi \Lambda_*^2} \sim H_{\rm c1}
\ee
A laboratory superconductor can only be penetrated
by an imposed magnetic field between the critical values \citep{tilley}.
As we have already mentioned, the neutron star problem is different. In this case, the
vortex state is accessible also for
$ B < H_{\rm c1} $, although it is metastable in this region. Hence, the transition
at $H_\mathrm{c1}$ is not likely to be as important for astrophysical systems.
In addition,
it is straightforward to show that type II superconductivity should be relevant for neutron stars.
By combining the expressions for the coherence and penetration lengths we find,
\be
\kappa_s \approx 0.4 \left ( \frac{m_\rp}{m_\rp^*} \right )^{3/2}
\left ( \frac{x_\rp}{0.1} \right )^{5/6} \rho_{14}^{5/6} \left (\frac{10^9~\mbox{K}}{T_{\rm cp}} \right )
\ee
For typical parameters pertaining to the outer core of a mature neutron star we have $\kappa_s < 1 $, i.e.
type II superconductivity should prevail \citep{baym}. However, at the critical density
\be
\rho_{\rm crit} \approx 3\times 10^{14}  \left ( \frac{m_\rp^*}{m_\rp} \right )^{9/5} \left (\frac{x_\rp}{0.1} \right )
\left (\frac{T_{\rm cp}}{10^9~\mbox{K}} \right )^{6/5}\, {\rm g}\, {\rm cm}^{-3}
\ee
beyond which $\kappa_s $ exceeds unity,  a type II $\rightarrow $ I phase transition should take place. This transition ought to
be relevant for neutron stars, since the critical density can easily be reached in the inner part of the core,
see Figure~\ref{fig:type}.

In a type I region the formation of proton vortices is energetically prohibited. Instead, the magnetic field is distributed
into regions of normal protons with large spatial dimensions (compared to the fluxtube cross-section)
and without a universal shape (see for example, \citet{sed05}).
The actual shape of these regions is mainly determined by the magnetic field distribution when the protons condense
(at the critical temperature $T_{\rm cp} $).  The lack of well-defined structures (like vortices) in a type I superconducting region
is the main obstacle preventing the formulation of macroscopic MHD for this region. For this reason,
we will not consider type I superconductivity further in this paper. However, since a portion of a neutron
star core may contain type I matter, future work needs to address this problem (see \citet{jones06} for an initial discussion).

So far, our survey of neutron star superconductivity has been somewhat idealised, and our development
of the MHD equations will be at a similar level. This is natural, since this is our first
investigation of the problem. However, it is useful to mention some other aspects
of superconductivity which could be relevant (and perhaps of key importance) for more realistic neutron star models.

One such aspect is the variation with density of the key physical parameters, like the length scales
$\xi_\rp$ and $\Lambda_*$, the critical temperature $T_{\rm cp} $ and the ratio $\kappa_s $. Using the relation
$ \Delta_\rp \approx 1.75 k_{\rm B} T_{\rm cp}$ between $T_{\rm cp} $ and the proton pairing energy gap
$\Delta_\rp $ and the approximate gaps discussed by \cite{viscous}
we can produce neutron star ``superconductivity cross-sections'' like that shown in Fig.~\ref{fig:type}.
This figure spans a typical range between ``strong'' and ``weak'' proton pairing (the chosen models correspond to the pairing cases ``e'' and ``f'' 
in \citet{viscous}, right and left panels, respectively),  
and illustrates the layered profile of proton  superconductivity in a neutron star core composed of npe matter only.

The results in Fig.~\ref{fig:type} show that an attempt to model realistic neutron stars would have to account for the
coexistence of type I, type II and normal regions and the detailed physics at the interfaces separating these regions.
In addition, the figure shows that the critical field $H_{\rm c2} $, as given by eqn.~(\ref{Hc2}),
can vary by up to an order of magnitude. While it may
reach a maximum $\sim 10^{16}\, {\rm G} $ it can drop to $\sim 10^{15}\, {\rm G} $ as the transition density is approached.


New features may also appear due to the coexistence of several mutually interacting superfluid and superconducting particle species.
This possibility has usually been
ignored in discussions of neutron star physics, for example,
in the calculation of the structure of a vortex. Yet, it may turn out that these interactions are important.
For the ``simple'' two-component mixture of superconducting protons and superfluid neutrons recent
work by \citet{alford08} suggests that the type II $\rightarrow $ I phase transition is significantly affected by the presence of the
neutron superfluid. The coupling between the two condensates results in a shift in the value of $\kappa_s $ at which the transition
takes place. Moreover, the coupling leads to the presence of a region populated by ``higher order''
proton vortices, each carrying a larger number of
flux quanta. This result hints at the complexity of the real type II $\rightarrow$ I transition and the presence of
an interface that could have impact on neutron star dynamics.

The departure from standard superconductivity could be even more pronounced in the inner core, especially
if it is populated by exotic particles. The most likely scenario allows for the presence
of hyperons in superfluid/superconducting states (for a recent review see \citet{schaffner08}). 
If we were to consider only the $\Sigma^{-} $ and $\Lambda$
hyperons, which are expected to appear first \citep{haensel}, then we
end up with a five-fluid system, containing two superfluids ($\rn, \Lambda $) two superconductors ($\rp, \Sigma^{-}$) and the normal
electrons. This is clearly a much more complicated multifluid system than that for the two-fluid outer core.
Moreover, hyperon cores may exhibit the phenomenology of so-called ``two-gap'' superconductors.
An example of such a system is liquid metallic hydrogen which exhibits a variety of vortex states, not observed in standard
single-component superconductors (see, for example, \citet{babaev}). This means that the high density part of the
superconductivity cross-sections shown in Fig.~\ref{fig:type} could be drastically different in neutron stars with hyperon cores.

Another issue that has not yet been discussed in detail is the character of superconductivity in magnetars.
For magnetars, the exterior dipole magnetic
field is estimated to be $\sim 10^{15}\, \mathrm{G} $ \citep{woods}. The strength of the interior field is unknown but it should be
comparable, or even higher, than the exterior one, probably with a mixed poloidal/toroidal topology \citep{braithwaite}.
This is interesting because
magnetic fields in the range $\sim 10^{15}- 10^{16}\, \mbox{G} $ may be strong enough to
suppress proton superconductivity (above $H_{\mathrm{c2}}$). Hence, it is not unreasonable to suggest that large portions of a 
magnetar core may not be superconducting. Another likely (and interesting!) possibility is that the interior magnetic field lies 
between the critical fields $H_{\rm c1} $ and $H_{\rm c2} $. Then the proton vortices will be so densely packed, in the sense
that their spacing  $d_\rp \lesssim \Lambda_* $, that they will interact. This regime of superconductivity
has not received any attention at all in the context of neutron stars, but it is known from laboratory systems that mutual vortex 
interactions could lead to a new type of ultra-low frequency magnetic oscillations \citep{gennes}.


\section{Multifluid MHD without vortices}
\label{sec:multi}

As a first step towards a formalism for mixed superfluids/superconductors, we discuss the derivation
of the MHD equations for a self-gravitating multifluid system composed of a number of charged and neutral
constituents (labeled by the indices $\{ \rx,\ry\}$) coupled to an electromagnetic field. At this point,
any number of constituents can be in a superfluid/superconducting state. Initially, we will not account for
the presence of vortices. This will be done later, in Section~\ref{sec:fullMHD}.

The discussion in this Section can be kept at an abstract level without the need to identify the various constituents.
Of course, at the end of the day, the resulting MHD equations will be specialised for modelling the outer core of a neutron star
composed of superfluid neutrons, superconducting protons and normal electrons.
It is worth noting that the model we will discuss, qhich is based on \citet{prix04}, represents the Newtonian limit of the relativistic model
developed by \citet{bc89}, see \citet{livrev} for a review. By including charged constituents, it is also the natural extension of the
flux-conservative model discussed by \citet{monster}. However, we will not attempt to develop an
analogous, generic, model for the dissipative problem. In fact, it is well-known that the development
of such a model, accounting for the electromagnetic field, is a real challenge.

The basic variables of the multifluid system are the  number densities $n_\rx$, the kinematic velocities
$v^i_\rx$, the mass $m_\rx $ and the charge $q_\rx $ carried by each particle of the  $\rx $ constituent.
Given these, we can define mass and charge densities and the associated fluxes/currents according to,
\bear
&& \rho_\rx = m_\rx n_\rx, \qquad     n_\rx^i = n_\rx v^i_\rx
\\
&&\sigma_\rx = q_\rx n_\rx, \qquad j_\rx^i = \sigma_\rx v^i_\rx
\eear
Summing over all components we get the total charge current, mass density and charge density,
\be
j^i = \sum_\rx j^i_\rx, \qquad \rho = \sum_\rx \rho_\rx, \qquad \sigma = \sum_\rx \sigma_\rx
\ee
The system can be described by a Lagrangian $\mathcal{L}$, which is a functional  $\cL (n_\rx,n^i_\rx,A_0,A^i,\Phi)$, where the
gravitational potential is $\Phi $, and   $A_0$ and $A^i $ represent the electromagnetic
degrees of freedom (the vector potential). The latter parameters are related to the electric field $E_i$ and
the magnetic field $B_i$
in the usual way, i.e., we have
\be
E_i = \nabla_i A_0 - \frac{1}{c} \partial_t A_i, \qquad B^i = \epsilon^{ijk} \nabla_j A_k
\ee
As a result, the two Maxwell equations
\be
\nabla_i B^i=0, \qquad \epsilon^{ijk} \nabla_j E_k = -\frac{1}{c} \partial_t B^i
\label{faraday}
\ee
are trivial identities.

Following \citet{prix05}, we assume a Lagrangian with minimal coupling. This means that the
``hydrodynamical'' contribution can be separated from the electromagnetic one. In other words, we
take as our Lagrangian
\be
\mathcal{L} = \mathcal{L}_\mathrm{H}( n_\rx, n_\rx^i)  + \mathcal{L}_\mathrm{EM} (E_i, B_i) + \mathcal{L}_\mathrm{grav} +
\left( \sigma A_0 + { 1 \over c} j_i A^i \right) - \rho \Phi
\ee
The electromagnetic and gravitational contributions to the Lagrangian are given by the usual expressions
\be
\mathcal{L}_\mathrm{EM} = { 1 \over 8 \pi} ( E^2 - B^2), \qquad 
\cL_{\rm grav} = -\frac{1}{8\pi G} (\nabla \Phi)^2
\ee
Meanwhile, the hydrodynamical piece is given by \citep{prix04}
\be
\mathcal{L}_\mathrm{H} = \sum_\rx { m_\rx n^\rx_i n_\rx^i \over 2 n_\rx} - \mathcal{E}
\ee
where the energy $\cE$ represents the ``equation of state''. In general this energy has a
functional form $\cE=\cE(n_\rx, w^2_\rxy)$, where $w^i_\rxy = v^i_\rx - v^i_\ry$ is the relative velocity 
between the $\rx$ and $\ry$ constituents. 

The remaining two Maxwell equations are obtained from the variation of $\mathcal{L}$ with respect to $A_0$ and $A^i$. This leads to
\be
\nabla_i D^i = 4\pi\sigma, \qquad
\epsilon^{ijk} \nabla_j H_k = \frac{4\pi}{c} j^i + \frac{1}{c} \partial_t D^i
\label{Max1}
\ee
where the ``dynamical'' fields $(D^i,H^i) $ are related to the ``kinematical'' fields $(E^i,B^i) $
according to
\be
D^i = 4\pi \frac{\partial \mathcal{L}}{\partial E_i}, \qquad  H^i = - 4\pi  \frac{\partial \mathcal{L}}{\partial B_i}
\ee
Similarly, variation of $\mathcal{L}$ with respect to the gravitational potential $\Phi$ leads to the
Poisson equation
\be
\nabla^2 \Phi = 4\pi G\rho
\ee
The hydrodynamical equations of motion are derived by varying $\mathcal{L}$ with respect to the basic
fluid variables $n_\rx$ and $n^i_\rx$ (see \citet{prix04} for details). If the total particle number for each constituent is assumed
fixed then one arrives at the particle and charge continuity equations,
\be
\partial_t n_\rx + \nabla_i n^i_\rx = 0, \qquad \partial_t \sigma_\rx + \nabla_i j_\rx^i = 0
\ee
For an isolated system, the conservation of momentum is then expressed by a collective Euler-type equation
\be
\sum_\rx \Bigg \{ n_\rx \partial_t \pi^i_\rx - n_\rx \nabla^i \pi_0^\rx -\epsilon^{ijk} n_j^\rx \epsilon_{klm}
\nabla^l \pi^m_\rx  \Bigg \} =0
\label{momeqns}
\ee
This equation features the canonical momenta
\be
\pi^\rx_i = \frac{\partial \cL}{\partial n_\rx^i} = p_i^\rx + \frac{q_\rx}{c} A_i
\ee
where we have introduced the hydrodynamical momenta \citep{prix04},
\be
p_i^\rx =  \frac{\partial \cL_{\rm H}}{\partial n_\rx^i}
= m_\rx \left ( v_i^\rx + \sum_\ry \varepsilon_{\rxy} w_i^\ryx \right )
\ee
This expression demonstrates the well known fact that the entrainment effect between the different particle species,
entering the formalism through the parameters
\be
\varepsilon_\rxy \equiv \frac{2}{\rho_\rx} \left ( \frac{\partial \mathcal{E}}{\partial w^2_\rxy} \right )_{n_\rx}
\label{entrain}
\ee
can misalign the momentum $p^i_\rx$ with respect to the corresponding flux $n^i_\rx$.

The scalar function $\pi^\rx_0 = \partial \cL/\partial n_\rx$ in the equations of motion is equal
to
\be
\pi^\rx_0 = -\mu_\rx -\frac{1}{2} m_\rx v^2_\rx + q_\rx A_0 - m_\rx \Phi
\label{pi0}
\ee
where
\be
\mu_\rx = \left( \frac{\partial \mathcal{E}}{\partial n_\rx} \right)_{n_\ry, w_\rxy^2}
\label{chemicals}
\ee
are the usual chemical potentials.

The momentum equation (\ref{momeqns}) can be re-written on an equivalent ``force-balance'' form (note that the signs
here are convention)
\be
\sum_\rx \Bigg \{\,  F^i_{\rm H x} - F_{\rm EM x}^i  + \rho_\rx \nabla^i \Phi  \, \Bigg \} = 0
\label{eom}
\ee
If there are additional (external) forces acting on the system they will appear on the
right-hand-side of this equation. The  ``hydrodynamical'' and electromagnetic force densities in
(\ref{eom}) are given by \citep{prix05},
\be
F_{\rm H x}^i = n_\rx \Bigg \{  (\partial_t + v^j_\rx \nabla_j ) p_\rx^i +  \nabla^i \mu_\rx
 + m_\rx \sum_\ry \left ( \varepsilon_{\rxy} w_j^\ryx \right ) \nabla^i v_\rx^j \Bigg \}
\label{fh}
\ee
\be
 F_{\rm EM x}^i = \sigma_\rx \left (\, E^i + \frac{1}{c} \epsilon^{ijk} v_j^\rx B_k  \, \right )
\label{fem}
\ee
At this point it is worth noting that in this model the number of Euler equation-type terms in (\ref{momeqns}) or (\ref{eom})
is the same as the number of constituents. In this sense, each constituent can be identified
with a ``fluid''. This identification can always be made in principle, but in many situations it is not appropriate.
If two or more constituents are comoving (i.e. have same $v^i_\rx $) then it is obviously only their ``conglomerate''
which represents an identifiable fluid degree of freedom. This is the case when the dynamical timescale is much longer than 
the relaxation due to interparticle scattering or electromagnetic coupling. Superfluidity suppresses
dissipative scattering and allows the components to have distinct dynamics.
 It is also worth pointing out that a single ``constituent''
fluid may have multiple dynamical degrees of freedom. This is the case for, for example, ${}^4 {\rm He}$ at low
temperatures, where a superfluid condensate coexists with a gas of thermal excitations (phonons and rotons). The
description of this system also requires two-fluid hydrodynamics (see \citet{nahelium} for a recent discussion).

The model we have discussed is general in the sense that it applies to any multifluid system interacting with an 
electromagnetic field. When compared to the standard equations of plasma physics \citep{jackson},
the formalism outlined here shows certain similarities but also significant differences.
The main difference is due to the presence of entrainment-induced couplings. This is known to be
a generic property of interacting Fermi liquids \citep{bashkin}.
Since one can associate the entrainment coefficient with an ``effective'' mass for each constituent, it is also
a familiar effect from other mixing-problems (like gas bubbles moving in liquids, see \citet{geurst}).
Additional differences will emerge when constituents become superfluid/superconducting and  the system
is endowed with rotational and magnetic vortices.


\subsection{The MHD approximation}
\label{sec:MHD}

The MHD approximation follows the assumption of macroscopic charge neutrality. This obviously
requires the various charged constituents to be carriers of opposite charge, and there must be an overall balance.
As discussed in many standard textbooks (cf. \citet{jackson,plasma}), charge neutrality holds to high precision at lengthscales
larger than the so-called Debye length and for time intervals larger than the plasma period ($\sim 10^{-21}$~s).
Typically, these requirements are easily met when charged particles (like electrons) are sufficiently mobile
that they neutralise any local charge. This renders the system oblivious to its true plasma
origin. In addition to charge neutrality,  the MHD approximation uses the simplifying assumption
of ``slow'' fluid motion (compared to the speed of light).
It is easy to verify that, for the conditions in neutron star interiors, the MHD approximation
is justified at any scale relevant to hydrodynamics \citep{easson79}.

Hereafter we will focus on a particular multifluid system: a three-constituent mixture of neutrons, protons and
electrons, with respective labels $\rx = \{\rn,\rp,\re\} $. This is the simplest model for the outer core of a neutron
star. Generalisations to other situations would be, at least in principle, straightforward. 
Consider, for example, the inclusion of muons. 
Then, due to the short timescale for electron-muon scattering \citep{mendell}, the electrons and muons will rapidly relax to be 
co-moving on average. It is thus easy to extend our model to cover this case, but we have opted not to do this 
(in order to keep the discussion clear). We refer the interested reader to \citet{mendell} for details and briefly remark, where appropriate, 
how our results would change if muons were present.

Taking the charges to be $q_\rp=-q_\re=e$ and $q_\rn=0$, and
imposing overall neutrality $\sigma = \sigma_\re + \sigma_\rp = 0 $ we must have  $\nabla_i j^i = 0$ and,
\be
n_\re = n_\rp  \quad \Rightarrow \quad j^i =  e n_\rp w^i_\rpe
\label{neutral}
\ee
The slow motion approximation then allows the elimination of the displacement current from (\ref{Max1}). This leads to Amp\'ere's law,
\be
j^i = \frac{c}{4\pi}\epsilon^{ijk} \nabla_j H_k
\label{ampere}
\ee
Let us now consider the momentum equations for this system.
Without any loss of generality, equation (\ref{eom}) can be separated by prescribing a
coupling force $ F^i_{\rm mf} $ between the neutrons and the combined charged components.
At this point this force is unspecified, but we will later identify it with
the vortex-mediated mutual friction force. This leads to the system
\be
n_\rn \Bigg \{ \,  (\partial_t + v^j_\rn \nabla_j )\, p_\rn^i + \nabla^i \mu_\rn
+ m_\rn \en w_{\rm pn}^j \nabla^i v_j^\rn + m_\rn \nabla^i \Phi   \, \Bigg \} = F^i_{\rm mf}
\label{eulern0}
\ee
\be
n_\rp \Bigg \{ \,  (\partial_t + v^j_\rp \nabla_j )\, p_\rp^i  + \nabla^i \mu_\rp
- m_\rp \ep w_{\rm pn}^j \nabla^i v_j^\rp \, \Bigg \}  
+ \rho_\re (\partial_t + v^j_\re \nabla_j)\, v^i_\re + n_\re \nabla^i \mu_\re
+ (\rho_\rp + \rho_\re) \nabla^i \Phi
= -F^i_{\rm mf} + F^i_{\rm L}
\label{p+e}
\ee
where
\be
F^i_{\rm L} =  \frac{1}{c} \epsilon^{ilm} j_l B_m
\label{Lordef}
\ee
is the standard Lorentz force density.

In equations (\ref{eulern0})--(\ref{p+e}) we have only considered the entrainment between the neutrons and the protons.
 The coefficients $\en = \varepsilon_\rnp $ and $\ep = \varepsilon_\rpn $ are related
 since $\rho_\rn \en = \ep \rho_\rp $, c.f. \eqref{entrain}.
We assume a vanishing entrainment for the electrons, i.e., let $\varepsilon_{\rm ex} =0 $,
in order to simplify the problem. There is no compelling physical argument \underline{not} to make this assumption.
The situation is different for the entrainment between the baryons, which is due to the strong interaction.

Even though the system contains three distinct constituents, it is possible
to ``reduce'' it to an effective two-fluid system by exploiting the smallness of the electron inertia
(compared to that of protons and neutrons). Mathematically, this amounts to neglecting the electron inertia
in (\ref{p+e}). Thus, we obtain
\be
\rho_\rp \Bigg \{ (\partial_t + v^j_\rp \nabla_j )\, (v^i_\rp - \ep w^i_{\rm pn} ) + \nabla^i \left (  \tilde{\mu} + \Phi \right )
- \ep w_{\rm pn}^j \nabla^i v_j^\rp \Bigg \}  =  F^i_{\rm L}  - F^i_{\rm mf}
\label{2to1}
\ee
where we have introduced the combined proton-electron chemical potential,
\be
\mut = \frac{\mu_\rp + \mu_\re}{m_\rp}
\ee
In the following we will use $m=m_\rn=m_\rp$. We arrived at (\ref{2to1}) by neglecting terms which are a factor $ m_\re/m\ \ll 1 $
smaller than those we retain. In other words, these equations are accurate\footnote{In the presence of muons eqn.~(\ref{2to1}) is still 
valid provided we redefine eqn. (38) according to
$$
\rho_\rp \nabla_i \tilde{\mu} \approx (\rho_\rp + \rho_\re + \rho_\mu) \nabla_i \tilde{\mu} = \rho_\rp
\nabla_i \tilde{\mu}_\rp + \rho_\re \nabla_i \tilde{\mu}_\re + \rho_\mu \nabla_i \tilde{\mu}_\mu
$$ 
The equations are in this case accurate up to corrections of the order of $(\rho_\re + \rho_\mu)/\rho_p \sim 10^{-2}$ if we assume that 
$n_\mu\lesssim 0.1n_\rp$ which is reasonable in the outer core of a neutron star.} up to about one part in $10^3$.
The resulting Euler equation (\ref{2to1}) is effectively a single-fluid equation,
without explicit dependence on the electron velocity. The electron velocity enters only the
total current $j^i$ which, however, can be eliminated by means of (\ref{ampere}). If we ignore the Lorentz force,
e.g. if the protons and electrons are considered to be comoving,  then
the pair of equations \eqref{eulern0} and \eqref{2to1} reduce to the non-magnetic two-fluid model that has
been used in a number of studies of superfluid neutron stars dynamics 
(see for example \citet{prix99,rieutord,na01,chamel06,prec1,glitch,andrea}).


\subsection{The induction equation}
\label{sec:induct}

In order to complete the set of MHD equations, we need to provide  a relation between the electric field and the fluid variables.
Given such a relation, we can  eliminate the electric field from  Faraday's law (\ref{faraday}) and derive
a final equation relating the magnetic field to the fluid variables, the so-called induction equation.

We take the Euler equation for the electron fluid as our starting  point.
This equation can be separated  from the collective equation of motion (\ref{eom}) by introducing (at this point unspecified)
coupling forces $F^i_\rpe$ and $F^i_\rne $ between the electrons and the proton and neutron fluids, respectively.
This means that we have
\be
\rho_\re \left [\, ( \partial_t + v^j_\re \nabla_j ) v_\re^i + \nabla^i \left ( \tilde{\mu}_\re
 + \Phi \right ) \, \right ] + e n_\re \left ( E^i + \frac{1}{c}\epsilon^{ijk} v_j^\re B_k  \right )
= - \left  (\, F^i_{\rm ne} + F^i_{\rm pe} \, \right )
\label{eulere}
\ee
Again ignoring  the electron inertia we obtain an expression for the electric field
\be
E^i \approx -\frac{1}{c} \epsilon^{ijk} v_j^\re  B_k
-\frac{1}{e} \nabla^i \left ( \mu_\re + m_\re \Phi \right ) -\frac{1}{e n_\re}\left (\,  F^i_{\rm ne} + F^i_{\rm pe} \, \right )
\label{electric}
\ee
Inserting this result in (\ref{faraday}) and eliminating the current, we get
\be
\partial_t B^i = \epsilon^{ijk} \epsilon_{klm} \nabla_j  \left ( v^l_\re  B^m \right)
+ \frac{1}{a_\rp} \epsilon^{ijk} \nabla_j \left [ \frac{1}{\rho_\rp} \left (\, F_k^{\rm ne} + F_k^{\rm pe} \, \right )
\right ]
\label{ind}
\ee
where
\be
a_\rp = \frac{e}{m c}
\ee
The electron velocity can be eliminated via
\be
v^i_\re = v^i_\rp - \frac{1}{e n_\rp} j^i = v^i_\rp -\frac{1}{ 4\pi  a_\rp \rho_\rp}
\epsilon^{ijk} \nabla_j H_k
\label{ve}
\ee
Equation (\ref{ind}) is the general expression for the induction equation in our multifluid system\footnote{In fact, eqn.~(\ref{ind}) is valid 
even if muons are present, but the forces $F^i_{\rm ne}$ and $F^i_{\rm pe}$ should of course be reinterpreted as 
being due to interactions with the combined electron-muon fluid.}. 
It differs from the standard MHD induction equation, c.f. \cite{STbook}, in two ways. Most apparently,
the total coupling force $ F^i_\re \equiv F^i_{\rm pe} + F^i_{\rm ne}$ does not necessarily take the form of a frictional force
$F^i_\re \propto j^i$ (which in standard MHD would lead to an Ohmic dissipation term in (\ref{ind})). 
This point will become obvious when we discuss the nature of $F^i_\re$ in Section~\ref{sec:mf}.
In addition, we will later establish that $H^i$ (entering in \eqref{ind} implicitly through the use of \eqref{ve}) 
differs considerably from the standard result. In fact, in Section~\ref{sec:ampere} we will establish that 
$H^i = b_\mathrm{L}^i$, where $b_\mathrm{L}^i$ is the London field. Both these effects may lead to important differences regarding, 
for example, the magnetic field evolution.


\section{MHD in the presence of vortices}
\label{sec:fullMHD}

In order to model the conditions expected in the outer core of a mature neutron star, we
need a formalism that accounts for neutron superfluidity and proton superconductivity.
This is of key importance since  fundamental features like the
bulk rotation and the magnetic flux may be ``locked''  into a population of quantised neutron and proton
vortices. These need to be incorporated in the formalism.

The presence of Fermi-scale sized vortices leads to a physical system
with multi-scale dynamics. However, as in the case of superfluid Helium,  it is usually sufficient
to consider the smooth-averaged properties of the neutron and proton vortex arrays. This is at least the case as long as one is mainly interested in
large-scale dynamics. Averaging results in macroscopic quantisation conditions, the so-called Onsager-Feynman
relations (equations (\ref{L1}) and (\ref{L2}) below), which involve the fluid velocities and the magnetic field.
In the case of neutron stars it is obvious that one can only make progress by working at the averaged (macroscopic)
level\footnote{This is in contrast with laboratory systems like cold atom condensates which often require
an approach that ``resolves'' individual vortices.}. There is no
way that one will ever be able to study the dynamics of the immense number of individual vortices.
Neither would one want to, for the result would likely be too complex to interpret. The averaging is essential, but comes at a price. 
The averaged model is essentially phenomenological, and the link to
the key microscopic parameters is not always clear. In fact, one of the main challenges concerns the
inference of macroscopic parameters from detailed microscopic models.

In addition to the quantisation conditions, the presence of vortices leads to new forces in the
hydrodynamical equations. Loosely speaking,  these forces tend to be of two different kinds: (i) resistive
``drag'' forces resulting from the interaction between the vortices and the fluids, leading to
effective ``mutual friction'' coupling terms in the Euler equations, (ii) forces due to the intrinsic
``tension'' of the vortices themselves. So far, we have ignored all such forces.
The following Sections are devoted to the discussion of the forces and
the associated modification of the MHD equations from Section~\ref{sec:multi}.

Before proceeding to the detailed analysis, it is important to make a comment on notation. In the rest of
the paper, the various parameters entering the MHD formalism refer to averaged quantities 
and as such the adopted notation should formally distinguish them from the corresponding parameters of the preceding 
vortex-free formalism. However, in order to keep the presentation as simple as possible, we opt to use the
same notation. In doing so, there should be no danger of confusion since the remainder of the paper is focussed on
vortex-averaged MHD.


\subsection{The macroscopic quantisation conditions}
\label{sec:q_conditions}

Due to the fundamental quantum nature of the superfluid/superconducting condensates, neutron and proton vortices
will be quantised in such a way that they carry a single quantum of momentum circulation\footnote{It would be relatively straightforward
to extend our analysis to let each proton vortex carry a larger flux. However, we expect that the most common case will be that of 
single quantum vortices. This case should be energetically favoured in most situations.}.
We will assume that both neutron and proton vortices are locally arranged in rectilinear 
arrays\footnote{Although it is usually assumed that the vortices in a neutron star core are ``aligned'', there is no reason
why this should be so. In fact, from the many studies of turbulence in superfluid laboratory systems one may expect that the
generic situation involves vortex tangles. The implications of such tangles for neutron star dynamics is not yet well understood.
Nevertheless, it is clear from the first efforts aimed at exploring this problem \citep{peralta,turbulence} that turbulence could 
play a key role in many situations.}
directed along unit vectors $\hat{\kappa}^i_\rn $ and $\hat{\kappa}^i_\rp$ (we use a `hat' to denote unit vectors). 
The two arrays have surface densities $\cN_\rn $ and $\cN_\rp $, respectively. At the hydrodynamics level, we then have \citep{menlin,prix05},
\be
\cW^i_\rn = \frac{1}{m} \epsilon^{ijk} \nabla_j p^\rn_k = \cN_\rn \kappa_\rn^i
\label{L1}
\ee
and
\be
\cW^i_\rp = \frac{1}{m} \epsilon^{ijk} \nabla_j \pi^\rp_k =
\frac{1}{m} \epsilon^{ijk} \nabla_j  p^\rp_k +  a_\rp B^i = \cN_\rp \kappa_\rp^i
\label{L2}
\ee
We have used $\kappa_\rx^i = \kappa \hat{\kappa}^i_\rx $ with $\kappa = h / 2m$.
It is important to note that the quantised ``vorticities'' refer to the circulation of the (averaged) canonical momenta
$ \pi^i_\rx =  p^i_\rx + (q_\rx/c) A^i $ rather than the circulation of velocity. It is the canonical
momentum which is related to the gradient of each condensate's wavefunction phase $ \phi_\rx$,
leading to the Onsager-Feynman quantisation condition
\be
\oint   \pi^i_\rx dl_i= (\hbar/2) \oint  (\nabla^i \phi_\rx)  dl_i = h/2 \ .
\ee
The vortices interact with their environment mainly through their intrinsic magnetic fields. These fields are  of the order
of $10^{15}~\mbox{G} $ and  extend only a very short distance away from the vortex core (see Appendix).
The magnetic field of the neutron vortices is generated by the current of  protons that are entrained by the local
circulation associated with the vortex \citep{als}. The protons can react to flows on the very short lengthscale associated
with the vortex core since they themselves form a condensate with a short coherence length. The mean-free path of the electrons
is much larger, which means that they cannot respond collectively on this short lengthscale.
Inferring the mean-free path ($\lambda_\re$) from results for electron-electron scattering shear viscosity 
(with coefficient $\eta$), we have
\be
\lambda_\re \approx { \eta \over \rho_\re c} 
\ee
since the electrons are expected to be relativistic. Scaling to typical parameter values, we then have
\be
\lambda_\re \approx 4 \times 10^{-3} \left( { \rho \over 10^{15} \mathrm{g/cm}^3} \right) 
\left( { 10^9\ \mathrm{K} \over T} \right)^2 \left( { 0.1 \over x_\rp} \right)\ \mathrm{cm} 
\ee
This result implies that the electrons do not act as a ``fluid'' on the mesoscopic scale of the individual vortices. 
As a result, the entrainment between neutrons and protons leads to the neutron vortices being magnetized.

The total averaged magnetic field $ B^i $  in (\ref{L2}) is a sum of three distinct
parts: the averaged fields $B^i_\rx$ contributed by the neutron and proton vortex arrays and the so-called London field $b_{\rm L}^i$.
The London field is a fundamental property of a rotating superconductor \citep{tilley}. In contrast to a
neutral superfluid which mimics bulk rotation by forming a vortex array, a superconductor can rotate without forming vortices. 
As we will soon see, the proton fluxtubes themselves are, in fact, not related to macroscopic rotation
at all. This crucial difference between superfluids and superconductors is encoded in the Onsager-Feynman
equations (\ref{L1}) and (\ref{L2}).

The total averaged magnetic field can  be written
\be
B^i  = b_{\rm L}^i + B_\rp^i  + B_\rn^i 
\label{aveB}
\ee
The averaged vortex fields are expressed in terms of the densities $\cN_\rx$ and the magnetic flux associated with
a single vortex,
\be
 B_\rx^i  =   \cN_\rx  \phi_\rx \hat{\kappa}_\rx^i
\label{BvBf}
\ee
Each proton vortex carries a single quantum of magnetic flux
\be
\phi_\rp = \phi_0 = \frac{hc}{2e}
\ee
Meanwhile, the flux associated with a neutron vortex can vary continuously, as it is induced by the proton current that is
entrained by the neutron circulation. As discussed in the Appendix, each neutron vortex carries a flux
\citep{als}
\be
\phi_\rn = -\frac{\ep}{1-\en} \phi_0
\ee
where the minus sign appears because the magnetic field of a neutron vortex is antialigned with $\kappa^i_\rn$.
Although the fluxes $\phi_\rp$ and $\phi_\rn$ are comparable, the total magnetic field is entirely dominated by
the proton vortex contribution since $\cN_\rp$ exceeds $\cN_\rn$ by several orders of magnitude for 
typical neutron star parameters.

By combining (\ref{aveB}) with (\ref{L1}) and (\ref{L2}) we find that the London field is given by
\be
b_\mathrm{L}^i = - { 1 \over m a_\rp} \left( \epsilon^{ijk} \nabla_j p^\rp_k - { \ep \over 1 - \en} \epsilon^{ijk}
\nabla_j p^\rn_k \right)
\ee
If the entrainment is assumed uniform, this simplifies to
\be
b^i_{\rm L} = - {  \epstar \over  a_\rp } \epsilon^{ijk} \nabla_j v_k^\rp
\label{London}
\ee
where we have defined, for later convenience,
\be
\epstar = \frac{1-\en -\ep}{1-\en}
\label{gam}
\ee
It is useful to note that the entrainment parameter can be expressed in terms of the effective proton mass
$ m_\rp^*$. Since $\ep = 1 - m_\rp^*/m_\rp$ \citep{petal} we find that
\be
\epstar = { - \rho_\rp + \rho m_\rp^*/m_\rp \over \rho - 2 \rho_\rp  + \rho_\rp m_\rp^*/m_\rp} \approx   {m_\rp^* \over m_\rp}
\ee
This should be a good approximation in a neutron star core, where the proton fraction is small.

In the case of uniform proton rotation with angular frequency $\Omega_\rp^i$,
eqn. (\ref{London}) becomes
\be
b^i_\mathrm{L} = - { 2 \epstar \over a_\rp }  \Omega_\rp^i
\label{bL_uniform}
\ee
This result clearly reveals the link between the London field and the rotation of the proton superconductor.
Inserting typical numerical values in (\ref{bL_uniform}) shows that the London field is exceedingly weak compared to
the typical magnetic field strengths in neutron stars,
\be
b_\mathrm{L} \sim  { 2 \Omega_\rp \over a_\rp} \approx 0.01 \left({P \over 0.1\ \mathrm{s}}\right)^{-1} \ \mathrm{G}
\label{bL_num}
\ee


\subsection{A variational derivation}
\label{sec:variational}

The MHD equations from Section~\ref{sec:MHD} need to be modified in the presence of neutron and proton vortices.
The new set of equations must be cognizant of the  quantisation conditions (\ref{L1}) and (\ref{L2}) and of the
forces associated with the vortices on a mesoscopic scale. At the level of a single vortex a Magnus force is
present when there is relative motion with respect to the macroscopic fluid flow. A tension force appears when the vortex ``bends'', 
much like the tension in a piano wire. There is also a Lorentz force due to the interaction of the
intrinsic vortex magnetic field with the charged fluids.

Once the above pieces are combined in a consistent way, the end product is a
set of macroscopic Euler equations featuring \underline{averaged} force densities $t^i_\rx $ due to the vortices.
That is, we have
\be
\rho_\rn \Bigg \{ (\partial_t + v^j_\rn \nabla_j )\, (v^i_\rn + \en w^i_{\rm pn} )  + \nabla^i \left ( \tilde{\mu}_\rn
+ \Phi \right ) + \en w_{\rm pn}^j \nabla^i v_j^\rn \Bigg \}   =  F^i_{\rm mf} + t_\rn^i
\label{eulern}
\ee
and
\be
\rho_\rp \Bigg \{ (\partial_t + v^j_\rp \nabla_j )\, (v^i_\rp - \ep w^i_{\rm pn} ) + \nabla^i \left ( \tilde{\mu}
+ \Phi \right ) - \ep w_{\rm pn}^j \nabla^i v_j^\rp \Bigg \}   =  - F^i_{\rm mf} + F^i_{\rm L}
+ t_\rp^i
\label{eulerp}
\ee
The inclusion of these forces is, however, not the only necessary modification to the equations of motion. As we discuss
later, cf. Section~\ref{sec:lorentz}, there are also significant changes to Amp\'ere's law and
the Lorentz force  $F_{\rm L}^i$ (due to the averaged \underline{macroscopic} current $j^i$ interacting with the averaged 
\underline{total} magnetic field $B^i$) for a type II superconductor.

The ``complete'' equations of motion can be derived by a suitable extension of the Lagrangian framework described
in Section~\ref{sec:multi}. This approach is closely related to the original variational multifluid formalism of \citet{bk} (later
adopted by \citet{menlin,mendell,mendell98}) which is based on the use of the total energy density, and to the more recent
framework of \citet{cl95}. 

The presence of neutron and proton vortices in the multifluid system can be taken into account by adding an extra
piece $\cL_\rv$ to the total Lagrangian,
\be
\mathcal{L} = \mathcal{L}_\mathrm{H}  + \mathcal{L}_\mathrm{EM}  + \mathcal{L}_\mathrm{grav}
+ \cL_\rv + \left( \sigma A_0 + { 1 \over c} j_i A^i \right) - \rho \Phi
\label{Ltot1}
\ee
This vortex term is a functional $\cL_\rv = \cL_\rv (n_\rx, \cW^i_\rx )$ of the canonical momentum vorticities and the 
number densities. Its specific form is given by the total averaged vortex contribution to the energy, which is obtained
by simply taking the product of the energy per unit length $\cE_{\rm vx}$ of a single vortex and the corresponding vortex density
\be
\cL_\rv = -\sum_\rx  \cN_\rx \cE_{\rm vx}
\label{Lv}
\ee

The addition of $\cL_\rv$ is not the only required modification to the Lagrangian.
When we discussed the variational approach to the magnetised fluid problem in Section~\ref{sec:multi} we made a point of
separating the hydrodynamical and electromagnetic contributions to the Lagrangian. When vortices are present this separation 
is no longer appropriate. Since both neutron and proton vortices are magnetised,
the piece (\ref{Lv}) also contains the averaged energy due to the local magnetic fields.
We need to modify $\cL_{\rm EM}$ accordingly and account for the contribution of the ``free'' electromagnetic field, i.e.
the London field. Thus, we have
\be
\cL_{\rm EM} = \frac{1}{8\pi} \left ( E^2 - b^2_{\rm L} \right )
\ee
Since, as we will shortly see, the London field is itself a functional $b_{\rm L}^i (n_\rx,B^i,\cW^i_\rx)$ it is more natural
to consider the combination
\be
\cL_0 \equiv \cL_\rv -\frac{b^2_{\rm L}}{8\pi}
\ee
when taking the variation of $\cL$. Then the total
Lagrangian is\footnote{In terms of the energy density used in the formalism of \citet{menlin,mendell,mendell98} the
present modifications to the total Lagrangian amount to an energy expression
$ \mathcal{E}_0 =   b_{\rm L}^2/8\pi +   \sum_\rx \cN_\rx \mathcal{E}_{\rm vx} $,
omitting the contributions that do not play a role in the present discussion.}
\be
\cL = \cL_{\rm H} + \cL_{\rm grav} + \frac{E^2}{8\pi} + \left( \sigma A_0 + { 1 \over c} j_i A^i \right) - \rho \Phi
+ \cL_0   \equiv \cL_{\rm vf} + \cL_0
\label{Ltot2}
\ee
where $\cL_{\rm vf}$ represents all the pieces that retain the same form as in the vortex-free
system, see Section~\ref{sec:multi}. 
In particular, it is important to point out that the minimum coupling term in (\ref{Ltot2})
refers to the \underline{total averaged} electromagnetic field, as given by eqn.~(\ref{aveB}).

Before proceeding with the variational calculation it should  be noted that, in (\ref{Ltot1}) and (\ref{Ltot2}) we have not accounted for
contributions due to interactions between neighbouring proton and neutron vortices. Neutron vortex-vortex interactions originate
from the long-range radial profile of the neutron circulation. The macroscopic manifestation is an elastic response of the
vortex array to perturbations that do not involve vortex ``bending''. This effective elasticity leads to the so-called Tkachenko
waves. These ultra-low frequency waves have recently been observed in laboratory superfluid systems \citep{cod,mizu}. Whether they
are relevant in the context of neutron stars is still an open issue, see \citet{noro} for a recent discussion.
The main interaction between proton vortices, which is due to their short-range magnetic fields, could become important if the
vortices  ``touch'', in the sense that $ d_\rp \lesssim \Lambda_\star $. This is possible provided the magnetic field is sufficiently
strong to satisfy $ H_{\rm c1} < B < H_{\rm c2} $. This regime could be relevant for magnetars with superconducting cores, but we will not 
consider it further here.

Returning to the Lagrangian $\mathcal{L} $ and working out the variation we get
\be
\delta \cL = \delta \cL_{\rm vf}  -  \sum_\rx \left \{  \lambda^\rx_i \delta \cW^i_\rx + \zeta_\rx \delta n_\rx
\right \} + \frac{1}{4\pi} H_i \delta B^i
\label{Uvar}
\ee
with the definitions
\be
\lambda_i^\rx \equiv -\frac{\partial \mathcal{L}_0}{\partial \cW^i_\rx},
\qquad \zeta_\rx \equiv -\frac{\partial \mathcal{L}_0}{\partial n_\rx}
\label{zeta_def}
\ee
As before
\be
H_i =  4\pi \frac{\partial \cL}{\partial B^i} =  4\pi \frac{\partial \cL_0}{\partial B^i}
\label{Hdef}
\ee
since the only $B$-dependent piece of $\cL$ is $\cL_0$.

From \eqref{Uvar} we can see that the chemical potentials will have additional electromagnetic contributions such that
\be
\mu_\rx \rightarrow \mu_\rx + \zeta_\rx
\ee
In order to arrive at the specific form of the vortex forces in the equations of motion we need to express the variations
$\delta \cW^i_\rx$ in terms of the Lagrangian fluid displacement field $\xi^i_\rx$ \citep{prix04}.
The required relation follows from the conservation of vorticity, which at the local level requires
that\footnote{This is analogous to the local mass conservation law for each species, which takes the form
\citep{CFS78,kirsty}
$$
\Delta n_\rx = \delta n_\rx + \xi^j_\rx \nabla_j n_\rx = - n_\rx \left( \nabla_j \xi^j_\rx \right)\ .
$$}
\be
\Delta \cW^i_\rx = - \cW^i_\rx \left( \nabla_j \xi^j_\rx \right) \quad \longrightarrow \quad 
\delta  \cW^i_\rx =  \epsilon^{ijk} \epsilon_{klm} \nabla_j ( \xi^l_\rx \cW^m_\rx )
\ee
where $\Delta$ represents a Lagrangian variation.
Given this result, it is straightforward to show that the vorticity terms manifest
themselves as the forces
\be
t^\rx_i  = - \epsilon_{ijk} \epsilon^{k l m}  \cW_\rx^j   \nabla_l \lambda_m^\rx =
\cW_\rx^j (\nabla_j \lambda^\rx_i - \nabla_i \lambda^\rx_j)
\label{T1}
\ee
in the equations of motion. This agrees with the conclusions of \citet{mendell}.

It is relatively easy to see why the force should take this form. We can rewrite
the original fluid equation (\ref{fh}) as
\be
 f_{\rm H x}^i = n_\rx \partial_t p_\rx^i +  n_\rx \nabla^i \left( \mu_\rx - { m \over 2} v_\rx^2 + v_\rx^j
p_j^\rx \right) - n_\rx \epsilon^{ijk} v^\rx_j \epsilon_{klm}\nabla^l p_\rx^m
\ee
Recalling that the variational analysis \citep{prix04} used a Lagrangian
$\mathcal{L}_\mathrm{H}(n_\rx^i)$, where $n_\rx^i= n_\rx v_\rx^i$ are the individual fluxes, we rewrite the last term as
\be
- \epsilon^{ijk} \epsilon_{klm} n^\rx_j \nabla^l p_\rx^m
\ee
Comparing this to (\ref{T1}) we see that the latter has exactly the form that one might
expect from the variational analysis.

With the vortex terms $\zeta_\rx$ and $t^i_\rx$ accounted for, the canonical equations of motion (\ref{momeqns}) become
(after combining eqns.~(\ref{L2}), (\ref{eulern}), (\ref{eulerp}) and (\ref{T1})), 
\be
\sum_\rx \Bigg \{  n_\rx \left [ \partial_t \pi^i_\rx - \nabla^i (\pi_0^\rx -\zeta_\rx ) \right ]
-\epsilon^{ijk} \left \{ n_j^\rx + \epsilon_{j p s} \nabla^p \lambda^s_\rx \right \}
\epsilon_{klm} \nabla^l \pi^m_\rx  \Bigg \} =0
\label{momeqns2}
\ee
Note that the scalar function $\pi^\rx_0$ is now defined as $\pi^\rx_0 = \partial \cL_{\rm vf}/\partial n_\rx$ and, therefore, 
it is still given by (\ref{pi0}).

In order to proceed to the calculation of the explicit form of $\cL_0$ and the resulting vortex terms $t^i_\rx$ and $\zeta_\rx$
we need to ensure that $b^i_{\rm L}$ has the required functional form. This is easily achieved by combining the conditions
(\ref{L1}) and (\ref{L2}),
\be
b_{\rm L}^i = B^i - \frac{1}{\kappa} \sum_\rx \phi_\rx \cW_\rx^i
\ee
Using this result, together with (recall that $\cN_\rx = \cW_\rx /\kappa$)
\be
\mathcal{L}_0 =  - \frac{1}{\kappa} \sum_\rx \cW_\rx \cE_{\rm vx} -\frac{1}{8\pi}  b_{\rm L}^2
\ee
we find 
\be
\lambda_\rx^i =  \frac{\cE_{\rm vx}}{\kappa} \hat{\cW}^i_\rx - { \phi_\rx \over 4 \pi \kappa}  b_\mathrm{L}^i
= \frac{1}{4\pi a_\rp} \left (  H_{\rm vx} \hat{\cW}^i_\rx -\frac{\phi_\rx}{\phi_0}\, b^i_{\rm L} \right )
\ee
where in the last equation we have defined
\be 
 H_{\rm vx} \equiv { 4 \pi  a_\rp \cE_{\rm vx} \over \kappa } =  { 4 \pi  \cE_{\rm vx} \over \phi_0}
\label{Hvx_def}
\ee
We have also also used $\hat{\cW}^i_\rx = \hat{\kappa}^i_\rx$.
Then from eqns.~(\ref{T1}) and (\ref{zeta_def}) it follows that
\be
 t^\rx_i = { 1 \over 4 \pi  a_\rp} \cW_\rx^j \left[  \nabla_j (H_{\rm vx} \hat{\cW}^\rx_i) - \nabla_i (H_{\rm vx} \hat{\cW}^\rx_j)
- \nabla_j \left ( \frac{\phi_\rx}{\phi_0}    b^\mathrm{L}_i \right ) + \nabla_i \left ( \frac{\phi_\rx}{\phi_0}  b^\mathrm{L}_j \right) \right]
\label{txforces}
\ee
and
\be
\zeta_\rx = { m \over 4 \pi a_\rp} \left[ \sum_{\ry \neq \rx}   \frac{\partial H_{\rm vy}}{\partial \rho_\rx} \cW_\ry
+ {\partial \over \partial \rho_\rx} \left( \frac{\ep}{1-\en} \right) \cW^\rn_i b_\mathrm{L}^i  \right]
\label{zeta}
\ee
The parameters $H_{\rm vx}$ play the role of effective magnetic fields. In particular, from the definition and the
explicit results for the vortex energies (eqns.~(\ref{energn}) and (\ref{energp}) below) we can identify 
$ H_{\rm vp} = H_{\rm c1} $ and estimate $H_{\rm vn} \approx  2 \times 10^{15} \rho_{14} \ \mathrm{G} $.
These results suggest that $H_{\rm vn} / H_{\rm c1} \sim 10 $.

Our final results (\ref{txforces}) and (\ref{zeta}) for the vortex forces $t^i_\rx$ and potentials 
$\zeta_\rx$ do not agree with those obtained by \citet{mendell,mendell98}. The difference is, however,
limited to  terms which depend on the generally weak London field. As we have already pointed out,
based on the original discussion by \citet{carter}, the origin of this discrepancy can be traced back to an erroneous vortex 
energy contribution used by Mendell. Translated into our notation this energy $(\cE_{\rm vx})_{\rm M}$  is
written as
\be
(\cE_{\rm vx})_{\rm M} = \cE_{\rm vx} + \frac{\phi_\rx}{4\pi} \hat{\kappa}_i^\rx b^i_{\rm L} 
\label{Emen}
\ee
From a conceptual point of view, the appearance of a macroscopic variable like $b^i_{\rm L}$ in the 
expression for the mesoscopic vortex energy is suspicious. As we explicitly show 
in Appendix~\ref{app:carter} (generalising the original calculation of \citet{carter}) it is, indeed, 
incorrect. The correct expressions for the neutron and proton vortex energies are given by
eqns.~(\ref{Energ_vort})-(\ref{evortn}).

For an isolated neutron vortex the energy is well approximated by
(see eqn.~(\ref{evortn}))
\be
\cE_{\rm vn} \approx {\kappa^2 \over 4 \pi} { \rho_\rn  \over 1-\en}  \log \left( \frac{d_\rn}{\xi_\rn} \right)
\label{energn}
\ee
Recall that the required energy integral is cut off both at the inner edge (letting the hollow vortex core be represented
by the coherence length $\xi_\rn$) and at a distance where the assumption that the vortex is isolated breaks down
(taking this to be the vortex separation $d_\rn$). Similarly, the energy of a proton vortex is
\be
\cE_{\rm vp} \approx \frac{\kappa^2 \rho_\rp}{4\pi\epstar}  \log \left( \frac{\Lambda_*}{\xi_\rp} \right)
\label{energp}
\ee
where the energy integral has a cut-off at a distance comparable to the London penetration length $\Lambda_*$.


\subsection{Amp\'ere's law for a type II superconductor}
\label{sec:ampere}

In normal matter (or in vacuum) the dynamical field $H^i$ is typically identified with the magnetic field
itself, i.e. $H^i=B^i$. However, this is no longer true for a type II superconductor. From the definition
(\ref{Hdef}) we obtain the, possibly surprising, result
\be
H_i = -4\pi \frac{\partial \cL_0}{\partial B^i} \quad \Rightarrow \quad
H^i = b^i_{\rm L}
\label{H1}
\ee
As far as we are aware, \cite{carter} were the first to arrive at this result, pointing out that the usual identification
$H^i= B^i$ should not be expected to hold in type~II superconductors. In retrospect, this result makes perfect
sense given that  the bulk of the magnetic field is carried by the quantized proton fluxtubes. 
The system is analogous to a magnetized medium rather than a magnetic classical fluid. 

It follows that Amp\'ere's law for a type II superconductor is now given by (neglecting friction forces)
\be
\epsilon^{ijk} \nabla_j b_k^{\rm L} = \frac{4\pi}{c} j^i = \frac{4\pi e n_\rp}{c} w^i_\rpe
\label{ampereII}
\ee
This is obviously different from the standard result\footnote{Our form for Amp\'ere's law does not agree with
the corresponding result of \citet{mendell,mendell98} who makes the usual identification $H^i = B^i$, as a result
of using the erroneous vortex energy (\ref{Emen}).}.

The result (\ref{ampereII}) allows us to draw some general conclusions. First, the presence of the
weak London field $b_\mathrm{L}^i$ instead of the full field $B^i$ in \eqref{ampereII}  suggests that the total 
macroscopic (averaged) current will be significantly weaker than what we would expect in normal matter, given that
in neutron stars we would expect $b_{\rm L} \ll B$.

Second, in the
case where the proton fluid is uniformly rotating, and the entrainment parameter $\epstar$ is also assumed
uniform, the combination of (\ref{ampereII}) and (\ref{bL_uniform}) leads to the somewhat surprising
conclusion that the proton and electron fluids are comoving, $v^i_\rp = v^i_\re$. In the more realistic
scenario where $\epstar$ is allowed to vary inside the star (while preserving axisymmetry with respect to the
spin axis) we obtain
\be
w^i_\rpe \approx 2 \Lambda^2_* \Omega_\rp \frac{\partial_r \epstar}{\epstar} \hat{\varphi}^i
\ee
where $r$ is the cylindrical radial coordinate.


\subsection{The total  magnetic forces in the equations of motion}
\label{sec:lorentz}

We now return to the complete equations (\ref{eulern}) and (\ref{eulerp}) and write them as
\be
\rho_\rn \Bigg \{ (\partial_t + v^j_\rn \nabla_j )\, (v^i_\rn + \en w^i_{\rm pn} )  + \nabla^i \left ( \tilde{\mu}_\rn
+ \Phi \right ) + \en w_{\rm pn}^j \nabla^i v_j^\rn \Bigg \}   =  F^i_{\rm mf} + t_\rn^i
-\rho_\rn \nabla^i \tilde{\zeta}_\rn
\label{eulern1}
\ee
and
\be
\rho_\rp \Bigg \{ (\partial_t + v^j_\rp \nabla_j )\, (v^i_\rp - \ep w^i_{\rm pn} ) + \nabla^i \left ( \tilde{\mu}
+  \Phi \right ) - \ep w_{\rm pn}^j \nabla^i v_j^\rp \Bigg \}   =  - F^i_{\rm mf} + F^i_{\rm mag}
\label{eulerp1}
\ee
having defined $\tilde{\zeta}_\rx = \zeta_\rx /m$ and the total effective magnetic force
\be
F^i_{\rm mag} \equiv t^i_\rp + F^i_{\rm L} -\rho_\rp  \nabla^i \tilde{\zeta}_\rp
\ee
exerted on the charged fluids.

Using the definition of the Lorentz force
\be
F_{\rm L}^i =  \frac{1}{c} \epsilon^{ilm} j_l B_m  =
\frac{1}{4\pi} B^j \left (\, \nabla_j H^i - \nabla^i H_j  \, \right )
\label{lor1}
\ee
we make the identification (\ref{H1}) and obtain
\be
F_{\rm L}^i = \frac{1}{4\pi} B^j \left (\, \nabla_j b^i_{\rm L} - \nabla^i b_j^{\rm L} \, \right )
\label{lor2}
\ee
This result is obviously different from the familiar Lorentz force expression
\be
F^i_\mathrm{norm} = { 1\over 4\pi } \left [ B^j \nabla_j B^i -\frac{1}{2} \nabla^i (B_j B^j) \right ]
\label{normal}
\ee
which follows after setting $ H^i=B^i $. In normal matter we would, of course, have $F^i_{\rm mag} = F^i_{\rm norm}$.
However, for a type II superconductor the combined force $t^i_\rp - \rho_\rp \nabla^i \tilde{\zeta}_\rp$ due to the proton vortices
must be accounted for in the magnetic force. After a slight rearrangement of terms we can write $t^i_\rp$ as
\be
 t^\rp_i = { 1 \over 4 \pi  a_\rp} \left[  \cW_\rp^j \left \{  \nabla_j (H_{\rm c1} \hat{\cW}^\rp_i)
- \nabla_i (H_{\rm c1} \hat{\cW}^\rp_j)
\right \} -  \frac{1}{m}  \epsilon^{jlm} \nabla_l p_m^\rp \left(\nabla_j b^\mathrm{L}_i - \nabla_i b^\mathrm{L}_j \right) \right]
-F^i_{\rm L}
\ee
Comparing this result to (\ref{lor2}) we see that a part of the proton vortex force
exactly cancels the usual Lorentz force. That is, we are left with the force
\be
F^i_\mathrm{mag} = \frac{1}{4\pi a_\rp} \Bigg\{ \cW^j_\rp \left[ \nabla_j( H_{c1}  \hat{\cW}_\rp^i )
-\nabla^ i( H_{c1} \hat{\cW}_j^\rp) \right]
- \frac{1}{m}  \epsilon^{jlm} \nabla_l p_m^\rp\left ( \nabla_j b^i_{\rm L} -  \nabla^i b_j^{\rm L} \right )  \Bigg\}
-\rho_\rp  \nabla^i \tilde{\zeta}_\rp
\label{fmag}
\ee
entirely due to the proton vortices (plus a contribution in $\zeta_\rp$ due to the neutron vortices).

In retrospect, this result is
not too surprising since the global magnetic field (excluding the weak London field) is ``locked'' to
the proton vortices. Compared to the standard MHD result (\ref{normal}), this  effective magnetic force
exhibits significant differences. This is easy to see if we consider the special case of a non-rotating star
and neglect entrainment. Then we have $\epsilon^{ijk} \nabla_j v_k^\rx=0 $, and it follows 
that\footnote{This magnetic force was first derived by Easson \& Pethick more than thirty years ago \citep{easson}, 
for a single component type II superconductor  using a thermodynamical approach.}
\be
F_{\rm mag}^i  = \frac{1}{4\pi} B^j \left[ \nabla_j( H_{c1}  \hat{B}^i )
-\nabla^ i( H_{c1} \hat{B}_j)\, \right] -\frac{\rho_\rp}{4\pi} \nabla^i
\left ( B \frac{\partial H_{\rm c1}}{\partial \rho_\rp} \right ) 
= \frac{1}{4\pi} \Bigg [ B^j \nabla_j \left  ( H_{\rm c1} \hat{B}^i \right ) 
- \nabla^i \left ( \rho_\rp B \frac{\partial H_{\rm c1}}{\partial \rho_\rp} \right ) \Bigg ] 
\label{suFmag}
\ee
Just like the normal matter expression (\ref{normal}), the superconducting magnetic force contains
a ``tension'' and a ``pressure'' term. However, both terms scale with $B^i$ and $ H^i_{c1} = H_{c1} \hat{B}^i $,
instead of $B^i $ alone. Moreover, the superconducting force has an extra piece due to the density dependence
of the energy of the proton vortices.

The ratio of the two forces
\be
\frac{F_{\rm mag}}{F_\mathrm{norm}}\sim \frac{H_{\rm c1}}{B}
\ee
will be significant for most neutron stars. This result has been known for a long time, see for example \citet{easson},
and several of its repercussions have been discussed in the literature.
For example, as discussed by \cite{mendell98}, the local speed of Alfv\'en-type waves is modified according to
$ v^2_{\rm A} \rightarrow H_{\rm c1} B/4\pi\rho_\rp  $. But the overall change in the magnitude of the magnetic
force is not the only modification imparted by superconductivity. There are also important structural differences
between the two forces $ F^i_{\rm mag} $ and $ F^i_{\rm norm} $ due to the dependence
$ H_{\rm c1} = H_{\rm c1} (\rho_\rx) $. Furthermore, we should not forget the presence of the force
$t^i_\rn -\rho_\rn \nabla^i \tilde{\zeta}_\rn$ in the neutron Euler equation. This force could be
viewed as an effective magnetic force acting on the neutron fluid, as a result of the coupling due to entrainment.

In other words, one has to bear in mind that MHD for superconducting neutron stars involves much more than a simple
rescaling of the  Alfv\'en wave speed.


\subsection{A conservation of vorticity argument}
\label{sec:Hall}

The variational derivation of the equations of motion is perhaps the most general and elegant
treatment of the problem. After all, it only requires us to provide the form of the Lagrangian.
Nevertheless, we want to consider the problem from a more
intuitive (albeit less general) point of view. To do this we generalise the method of \citet{hall}, originally developed
in the context of two-fluid hydrodynamics for superfluid Helium. This approach provides a conceptually
different derivation of the Euler equations, based on the kinematics of a conserved number of vortices carrying the
entire fluid momentum vorticity. It also requires the input of the forces that determine the motion of a single isolated neutron or
proton vortex. Thus, consistency between the two derivations allows us to identify the
total conservative force exerted on a single neutron or proton vortex, \underline{without} any need to study the detailed mesoscopic
vortex-fluid interaction.

The starting point of the derivation is the Onsager-Feynman conditions (\ref{L1}) and (\ref{L2}) which
we can collectively write as
\be
\cW^i_\rx = \cN_\rx \kappa_\rx^i
\label{Lboth}
\ee
We also need to use the fact that the vortex number density is conserved, i.e. $\cN_\rx$ obeys a continuity equation of the form
\be
\partial_t \cN_\rx + \nabla_j \left ( \cN_\rx u^j_\rx\right ) =0
\ee
where $ u^i_\rx$ is the collective vortex velocity within a typical fluid element. In a sense, this relation
\underline{defines} the averaged vortex velocity.
Taking the time derivative of (\ref{Lboth}) we have
\be
\partial_t \cW^i_\rx = -\kappa_\rx^i \nabla_j ( \cN_\rx  u^j_\rx ) + \cN_\rx \partial_t \kappa^i_\rx
\ee
Reshuffling terms and using $\nabla_i \cW^i_\rx = \epsilon^{ijk} \nabla_i \nabla_j \pi_k^\rx = 0 $ we obtain
\be
\partial_t \cW^i_\rx = \nabla_j \left ( \cW^j_\rx u^i_\rx \right ) -  \nabla_j \left ( \cW^i_\rx u^j_\rx \right )
+ \cN_\rx \left ( \partial_t \kappa^i_\rx + u^j_\rx \nabla_j \kappa^i_\rx -\kappa^j_\rx \nabla_j u^i_\rx \right )
\label{Hall0}
\ee
In the language of differential geometry the motion of a single vortex is represented by
the Lie-transportation of the vector $\kappa^i_\rx $ (which designates the local vortex direction) by the $ u^i_\rx $ flow,
i.e.
\be
\partial_t \kappa^i_\rx + \cL_u \kappa^i_\rx =0
\label{Lie_kappa}
\ee
where the Lie derivative is defined in the usual way
\be
\cL_u  \kappa^i_\rx  = u^j \nabla_j  \kappa^i_\rx -  \kappa^j_\rx \nabla_j u^i
\ee
Then (\ref{Hall0}) reduces to
\be
\partial_t \cW^i_\rx + \epsilon^{ijk} \nabla_j \left ( \epsilon_{klm} \cW^l_\rx u^m_\rx \right ) =0
\ee
which states that the canonical vorticities $ \cW^i_\rx $ are locally conserved and advected by the $u^i_\rx $ flow. 
Writing this result as a total outer product we obtain
\be
\partial_t \pi^i_\rx -\epsilon^{ijk} u_j^\rx \epsilon_{klm} \nabla^l \pi^m_\rx = \nabla^i \Psi_\rx
\label{Halleqn}
\ee
where $\Psi_\rx$ are (so far unspecified) scalar potentials. This result coincides with the classic formula due to
\citet{hall}, see also \citet{trev}. It is important to emphasize that (\ref{Halleqn}) applies to
any condensed constituent with vortices, in the present case $\rx=\{\rn,\rp\}$.

Alternatively, we can write (\ref{Halleqn}) as
\be
n_\rx \partial_t \pi^i_\rx - \epsilon^{ijk} n_j^\rx \epsilon_{klm} \nabla^l \pi^m_\rx -n_\rx \nabla^i \Psi_\rx
= \cN_\rx \rho_\rx \epsilon^{ijk}  \kappa_j^\rx w_k^{\rm xv}
\label{Halleqn1}
\ee
where $ w^i_{\rm xv} = v^i_\rx - u^i_\rx$.
The functional form of the left-hand-side of this equation coincides with the \underline{vortex-free} equations of motion
(\ref{momeqns}) after a suitable identification of the potential $\Psi_\rx$. The right-hand side term
appears only in the presence of vortices. It is not difficult to trace its origin if we recall that the Magnus force
exerted on a vortex (per unit length) by the associated fluid is given by \citep{magnus}
\be
f^i_{\rm Mx} = -\rho_\rx \epsilon^{ijk} \kappa_j^\rx w^{\rm xv}_k
\ee
Thus, we can identify $-\cN_\rx f^i_{\rm Mx}$, the right-hand side of \eqref{Halleqn1}, as the averaged reaction force 
exerted on a fluid element by the vortex array.

At this point it is clear that we can relate the macroscopic hydrodynamics we have discussed so far with the
mesoscopic dynamics of individual vortices. This can be achieved by simply taking the force balance for a single
vortex (for simplicity assuming a negligible vortex inertia), solving it with respect to the Magnus force, and subsequently
inserting the result in (\ref{Halleqn1}).

We will assume that  the force balance
for a single neutron or proton vortex takes the form
\be
f^i_{\rm Mx} + f^i_{\rm emx} + f^i_{\rm sx}  =0
\label{vortbal}
\ee
where $f^i_{\rm emx}$ is a Lorentz-type force due to the interaction of the local vortex magnetic field and
the charged fluids, and $f^i_{\rm sx}$ represents ``tension'' force due to the self-interaction when the vortex is
bent. Additional ``mutual friction'' forces may enter (\ref{vortbal}), but these will be considered
separately in Section~\ref{sec:mf}.

The explicit form of the tension self-force, at leading order in the vortex curvature, is
\be
f^i_{\rm sx} = \cE_{\rm vx} \hat{\kappa}^j_\rx \nabla_j \hat{\kappa}^i_\rx
\label{self}
\ee
where $\cE_{\rm vx}$ is the total energy (per unit length) of a straight vortex.

Returning to the Euler equations (\ref{Halleqn1}) we eliminate the Magnus force using (\ref{vortbal}),
and expand the canonical momenta $\pi^i_\rx$ into their hydrodynamical and electromagnetic pieces. This leads to
\begin{multline}
n_\rx \left ( \partial_t + v^j_\rx \nabla_j \right ) p^i_\rx
+n_\rx \nabla^i \left (\frac{1}{2} m v^2_\rx + e A_0  -\Psi_\rx - p^j_\rx v_j^\rx   \right  )
+ \rho_\rx \sum_\ry \left ( \varepsilon_{\rm xy} w_j^{\rm yx} \right )
\nabla^i v_\rx^j
\\
- e n_\rx \left ( E^i + \frac{1}{c} \epsilon^{ijk} v_j^\rx B_k \right ) =  F^i_{\rm em x} + \cT_\rx^i,
\qquad (\rx, \ry = \{ \rn,\rp \})
\end{multline}
where we have defined the averaged vortex Lorentz force
\be
 F^i_{\rm em x} =  \cN_\rx f^i_{\rm emx} 
\ee
and the averaged vortex tension (recall that $\hat{\cW}^i_\rx = \hat{\kappa}^i_\rx$)
\be
\cT_\rx^i = \cN_\rx f^i_{\rm sx} \quad \to \quad \cT_\rx^i =  { H_{\rm vx} \over 4 \pi a_\rp } \cW_\rx^j \nabla_j \hat{\cW}_\rx^i
\ee
The explicit Euler equations for the neutron and the combined proton-electron fluids are
\be
n_\rn \left ( \partial_t + v^j_\rn \nabla_j \right ) p^i_\rn
+n_\rn \nabla^i \left (\frac{1}{2} m v^2_\rn -\Psi_\rn - p^j_\rn v_j^\rn   \right  )
+ \rho_\rn \en w_j^{\rm pn} \nabla^i v_\rn^j
= F^i_{\rm em n} + \cT_\rn^i
\label{Heulern}
\ee
and
\be
n_\rp \left ( \partial_t + v^j_\rp \nabla_j \right ) p^i_\rp
+n_\rp \nabla^i \left (\frac{1}{2} m v^2_\rp + q_\rx A_0 -\Psi_\rp - p^j_\rp v_j^\rp + \mu_\re \right  )
- \rho_\rp \ep w_j^{\rm pn} \nabla^i v_\rp^j 
= F^i_{\rm em p} + \cT_\rp^i + \frac{en_\rp}{c} \epsilon^{ijk} w_j^{\rm pe} B_k
\label{Heulerp}
\ee
These expressions should be compared to the equations of motion (\ref{eulern}) and (\ref{eulerp})
(in the limit $F^i_{\rm mf}=0$), in combination with the eqns.~(\ref{T1}) and (\ref{zeta}) for the forces $t^i_\rx$
and the potentials $\zeta_\rx$, and the definition of the Lorentz force (\ref{Lordef}).
This comparison shows that we can write
\be
t^i_\rx = \cN_\rx\Bigg \{ -\frac{1}{4\pi a_\rp }\epsilon^{ijk} \kappa_j^\rx \epsilon_{klm} \nabla^l
\left ( H_{\rm vx} \hat{\cW}^m_\rx - \frac{\phi_\rx}{\phi_0} b^m_{\rm L} \right )
\Bigg \}
\ee
The term inside the bracket represents the force $-f^i_{\rm M x}$ for a single vortex.
Therefore this term must be equal to $f^i_{\rm emx} + f^i_{\rm sx}$, as required by (\ref{vortbal}). 

If, for the moment, we assume uniform densities (which also implies uniform entrainment) we have
\be
t^i_\rx = \cN_\rx\Bigg \{ \frac{H_{\rm vx}}{4\pi a_\rp }\kappa^j_\rx \nabla_j \hat{\cW}^i_\rx
+ \frac{1}{c }\phi_\rx\epsilon^{ijk} \hat{\kappa}_j^\rx w_k^{\rm pe}
\Bigg \} = \cT^i_\rx + \cN_\rx \Bigg \{ \frac{e n_\rp}{c }\phi_\rx\epsilon^{ijk} \hat{\kappa}_j^\rx w_k^{\rm pe}  \Bigg \}
\label{id1}
\ee
and, after identifying the tension force (\ref{self}), we obtain
\be
f^i_{\rm em x} = \frac{e n_\rp}{c }\phi_\rx\epsilon^{ijk} \hat{\kappa}_j^\rx w_k^{\rm pe}
= - \frac{1}{c }\phi_\rx\epsilon^{ijk} j_j \hat{\kappa}_k^\rx
\label{id2}
\ee
The final step we need to take is to identify
\bear
&&\Psi_\rn = -\left ( \mu_\rn + \zeta_\rn +  m \Phi \right ) +  \frac{1}{2} m v^2_\rn - p^j_\rn v_j^\rn
\\
&&\Psi_\rp = -\left ( \mu_\rp  + \zeta_\rp +  m \Phi \right ) +  \frac{1}{2} m v^2_\rp + e A_0 - p^j_\rp v_j^\rp
\eear

The detailed identification of the various vortex forces participating in the fluid Euler equations also
offers a different perspective on the Lorentz force cancellation result we discussed earlier.   
First we decompose $F_{\rm L}$ as
\be
F^i_{\rm L} = F^i_{\rm b} + \sum_\rx \cN_\rx \Bigg \{
\frac{\phi_\rx}{c} \epsilon^{ijk} \hat{\kappa}_j^\rx j_k \Bigg \} = F^i_{\rm b} 
-\sum_\rx \cN_\rx f^i_{\rm emx}
\ee
where $F^i_{\rm b}$ is the Lorentz force due to the interaction between the total current and the London
field alone. Then (\ref{Heulerp}) becomes 
\begin{multline}
n_\rp \left ( \partial_t + v^j_\rp \nabla_j \right ) p^i_\rp
+ n_\rp \nabla^i \left (\mu_\rp + \mu_\re + \zeta_\rp + m\Phi \right  )
- \rho_\rp \ep w_j^{\rm pn} \nabla^i v_\rp^j =
\\
=F^i_{\rm b} -\sum_\rx \cN_\rx f^i_{\rm emx} +  \cT_\rp^i + \cN_\rp f^i_{\rm emp}
= F^i_{\rm b} +  \cT_\rp^i - \cN_\rn f^i_{\rm emn} 
\label{id3}
\end{multline}
That is, the Lorentz force cancellation (modulo the weak London component $F^i_{\rm b}$)
occurs due to the exact mutual cancellation of the terms representing the electromagnetic interaction
between the proton vortex array and the charged particles.

Eqns.~(\ref{id1}) and (\ref{id2}) provide the explicit mesoscopic forces acting on a single vortex (together with
the Magnus force). They were ``derived'' under the approximation of uniform fluid densities, the same approximation
that we would normally have made had we wished to derive these forces from first principles, that is, by a proper study 
of the local mesoscopic vortex dynamics. However, the procedure presented here allows us to move beyond the uniform density
approximation and obtain the  vortex forces to a higher precision. As before, this can be done by requiring
consistency between the intuitive Hall derivation and the more rigorous variational approach. We can then make 
the following identifications for the forces acting on a single vortex:
\be
f^i_{\rm em x} =  \frac{1}{4\pi}\epsilon^{ijk} \hat{\kappa}_j^\rx \epsilon_{klm} \nabla^l
\left ( \phi_\rx b^m_{\rm L} \right )
\label{id4}
\ee
and
\be
f^i_{\rm sx} =  -\epsilon^{ijk} \hat{\kappa}_j^\rx \epsilon_{klm} \nabla^l
\left ( \cE_{\rm vx} \hat{\kappa}^m_\rx \right )
\label{id5}
\ee
At the end of the day we have derived a set of explicit expressions for the conservative forces acting 
on a single neutron or proton vortex. These results will play a key role in the derivation of the mutual friction coupling forces
between the fluids. 


\section{Mutual friction coupling}
\label{sec:mf}

To complete the multifluid MHD equations, we need to account for dissipation and coupling
due to the vortices. This so-called ``mutual friction'' is a key damping agent for
relative fluid motion. It tends to be introduced phenomenologically \citep{als,mendell,sidery06}.
This is natural since the associated microphysics is complicated and not well understood.
 So far, we have included the mutual friction ``symbolically'' in the form of the  coupling
forces $F^i_{\rm mf},~F^i_\rpe$ and $F^i_\rne$ (see Section~\ref{sec:MHD}). We will now discuss
phenomenological expressions for these forces and provide the final set of MHD equations for a
mixed neutron superfluid/proton (type~II) superconductor.

The prescription for deriving the mutual friction force is well established. It was
first introduced in the classic work of \citet{HV} in the context of Helium superfluidity.
Initially, the mutual friction is expressed as a resistivity which depends
on the relative motion between a given neutron or proton vortex (moving with velocities
$u_\rn^i$ and $u^i_\rp$, respectively) and the various fluids. By considering the equation of motion 
for a single vortex (essentially a statement of force balance since the
inertia of a vortex is negligible) it is possible to express the $u_\rx^i$ velocities
in terms of the (smooth averaged) fluid velocities $v^i_\rx $. These relations are then
fed back into the original force expression, which now depends only on the fluid velocities
$v^i_\rx $ and phenomenological coupling coefficients.


\subsection{The mutual friction}

We will consider two principal friction mechanisms.
First of all, electrons can scatter dissipatively off of the vortex magnetic field \citep{als}.
The associated drag force (per unit vortex length) takes the form
 \be
f^i_{\rm D x} = R_\rx ( v^i_\re -u^i_\rx )
\label{drag1}
\ee
The coefficient $R_\rx$ follows from the relaxation time for electron flow through an
array of vortices \citep{als} and depends on the  magnetic fluxes $\phi_\rx$. 
For the neutron vortices, in particular, it will obviously depend on the strength of the entrainment.
In both cases the resistivity (\ref{drag1}) will tend
to relax the vortex motion to the electron flow.

In general, the  coefficients $ R_\rx$ depend on the proton and neutron number densities $ (n_\rp,n_\rn) $
and will vary considerably as we move from the crust-core interface to the inner core.
They are also sensitive (through their dependence on the size of the vortex cores)
to the finite ambient temperature $T$ of the neutron star core. This will be particularly important
in regions near the superfluid phase transition, where the
thermal energy is comparable to the neutron and proton pairing energy gaps, i.e. $ k_B T \sim \Delta_\rx $.
These complications have been ignored in many calculations, and the drag coefficients simply
taken to be uniform. This simplification is not really required in the derivation of the
mutual friction forces, but it makes it easier to include these forces in calculations of neutron star oscillation modes etcetera.

The second mutual friction mechanism that we consider is due to the direct (magnetic) interaction between neutron and
proton vortices. This interaction is presently only understood at a rudimentary level.
Simple back of the envelope estimates suggest that it could be strong enough to lead to neutron vortices ``pinning'' to the proton
vortices \citep{sauls,ruderman}. If the relative neutron-proton vortex velocity is high enough to prevent pinning,
the motion may still be efficiently damped as a result of the excitation of Kelvin waves propagating along the vortices \citep{link03}.

Given the lack of a detailed neutron-proton vortex interaction force, the best we can do is adopt a phenomenological approach.
Hence, we assume a resistive force of the usual form
\be
f^i_\rv = C_\rv  ( u^i_\rp -u^i_\rn )
\label{drag3}
\ee
where $C_\rv$ in another unspecified coefficient. This form for $f^i_\rv$ is undoubtedly a gross
simplification. However, the subsequent analysis will not be straightforward if the force takes a significantly
different form.  If, for example, $C_\rv$ were to have some intrinsic directionality
(perhaps in the form of a tensorial $C^{ij}_\rv $) the final mutual friction forces would be significantly
more complicated. One may, in fact, have to develop an entirely different strategy to deal with the problem.
Future work should address this issue. We need a derivation of the neutron-proton vortex interaction based on the
detailed meso-scale dynamics.

In principle, we should also consider the back-reaction force exerted by each neutron vortex
on the proton vortices. However, the force acting on an individual proton vortex will typically be much weaker than (\ref{drag3}),
and moreover, only a small fraction of the proton vortex population participates in the interaction.
Basically, this is due to the enormous difference in the density of the neutron and proton vortex arrays.
The separation between individual neutron vortices is roughly
\be
d_\rn \sim \cN_\rn^{-1/2} \sim 4\times 10^{-4} \left({P \over 1\ \mathrm{ms}}\right)^{1/2}\ \mathrm{cm}
\ee
where $P$ is the spin period. Meanwhile, the distance separating proton vortices can be approximated by
\be
d_\rp \sim \cN_\rp^{-1/2} \sim 5\times 10^{-10} \left({B\over 10^{12}\ \mathrm{G}}\right)^{-1/2}\ \mathrm{cm}
\ee
If we consider a fluid element of size $\sim L$ we would expect a number $ N_\rn \sim (L/d_\rn)^2 $
and  $ N_\rp \sim (L/d_\rp)^2 $ of neutron and proton vortices, respectively, within the element. Assuming a
number $ N_{\rm pin} \sim L/d_\rp $ of interaction sites per neutron vortex inside the element, we
see that
\be
N_\mathrm{pin} \sim 2\times 10^9  \left({B\over 10^{12}\ \mathrm{G}}\right)^{1/2} \left( {L \over 1\ \mathrm{cm}} \right)
\ee
Comparing this to the total number of interaction sites per proton vortex,
\be
N_{\rm tot}/N_\rp = N_{\rm pin} N_\rn/N_\rp \sim 3\times 10^{-3} \left({B\over 10^{12}\ \mathrm{G}}\right)^{-1/2}
\left({P \over 1\ \mathrm{ms}}\right)^{-1}
\left( {L \over 1\ \mathrm{cm}} \right)
\ee
we see that the backreaction onto the proton vortices can safely be ignored.


\subsection{Vortex kinematics}

As we have already discussed in Sections~4.2--3, the  motion of an individual vortex can be determined once all
forces acting on it have been identified. In addition to the resistivity
there are conservative forces present. One of these is the  Magnus force, which is present when a vortex moves relative to
a  ``background'' flow. We also need to account for the conservative electromagnetic interaction
between the magnetic field of a given vortex and the charged particles. Each vortex is also affected by a tension self-force,
see Section~\ref{sec:Hall}. However, as discussed in the same Section, if we assume that the vortex arrays are locally
straight we can extract the force acting on a single vortex from the hydrodynamical forces  (\ref{txforces})
simply by dividing by the relevant surface density $\cN_\rx$. This procedure leads to the vortex forces
(\ref{id4}) and (\ref{id5}).

Having identified the forces acting on each vortex we can write down the corresponding force balance equations.
In the case of a neutron vortex we have,
\be
f^i_{\rm Mn} +  f^i_{\rm emn}  + f^i_{\rm sn} + f^i_{\rm Dn} + f^i_\rv = 0
\label{eomv}
\ee
Meanwhile, for a proton vortex we get
\be
f^i_{\rm Mp} +  f^i_{\rm emp}  + f^i_{\rm sp} + f^i_{\rm Dp} = 0
\label{eomf}
\ee
In order to invert (\ref{eomf}) and obtain the proton vortex velocity $u^i_\rp$ we follow the strategy from the classic
work of \citet{HV}. Taking repeated cross products of (\ref{eomf}) with $\hat{\kappa}^i_\rp$  we arrive at
\be
u^i_\rp = v^i_\re + \frac{1}{1+ \cR_\rp^2} \Bigg ( \cR_\rp f^i_\star + \epsilon^{ijk} \hat{\kappa}_j^\rp f_{\star k} \Bigg )
\label{velf}
\ee
where
\be
f^i_\star = \epsilon^{ijk} \hat{\kappa}_j^\rp w^{\rm ep}_k + \frac{1}{\rho_\rp \kappa} \left (  f^i_{\rm emp}  + f^i_{\rm sp} \right )
\ee
We have also introduced the dimensionless resistivity
\be
\cR_\rp \equiv \frac{R_\rp}{\rho_\rp \kappa}
\ee
In order to obtain the neutron vortex velocity we first  insert (\ref{velf}) in (\ref{eomv}).
This leads to
\be
\epsilon^{ijk} \hat{\kappa}_j^\rn (u^\rn_k -v^\rn_k) +
\cR ( v^i_\re -u^i_\rn ) + D^i = 0
\label{eomv2}
\ee
where we have defined
\be
D^i = \frac{1}{\rho_\rn \kappa} \left( f^i_{\rm emn} + f^i_{\rm sn} \right )
 + \cC  \Bigg \{ \cR_\rp f^i_\star + \epsilon^{ijk} \hat{\kappa}_j^\rp f_{\star k} \Bigg \}
\ee
and
\be
\cC_\rv \equiv \frac{C_\rv}{\rho_\rn \kappa}\ ,     \qquad  \cR_\rn \equiv \frac{R_\rn}{\rho_\rn \kappa}
\quad \mbox{and} \quad
 \cC \equiv \frac{\cC_\rv}{1+ \cR_\rp^2} \ ,
\qquad \cR \equiv \cR_\rn + \cC_\rv
\ee
Basically, there are three dimensionless coefficients in the problem.

Repeating the previous inversion procedure we arrive at
\be
u^i_\rn = v^i_\re + \frac{1}{1 + \cR^2 } \Bigg (  \cR f^i +
\epsilon^{ijk} \hat{\kappa}^\rn_j f_{k} + \frac{1}{\cR} \hat{\kappa}_j^\rn D^j  \hat{\kappa}^i_\rn   \Bigg )
\label{velv}
\ee
where
\be
f^i = \epsilon^{ijk} \hat{\kappa}_j^\rn w^{\rm en}_k + D^i
\ee
According to  (\ref{velf}) and (\ref{velv}) we should, in general, expect relative motion between
the vortices and the fluids.


\subsection{The mutual friction forces}

The final step is to translate the forces acting on a single vortex into hydrodynamical
(smooth averaged) mutual friction forces. This is a straightforward operation if the vortices move collectively and the
respective arrays are locally straight (at the level of each fluid element). 
However, this is only the simplest situation (see the footnote in Section~\ref{sec:q_conditions}). 
Nevertheless, we make this assumption in order to make immediate progress. It means that we simply need to multiply each force expression by the
corresponding vortex surface density. The resulting forces (per unit volume) can be identified with the coupling forces
in the Euler equations (\ref{eulere}), (\ref{eulern}) and (\ref{eulerp}). This way, we arrive at
\bear
&& F^i_{\rm mf} = \cN_\rn \left ( f^i_{\rm Dn} + f^i_\rv \right )
= \rho_\rn \cW_\rn \left [  \cR_\rn (v^i_\re -u^i_\rn) + \cC_\rv (u^i_\rp -u^i_\rn) \right ]
\label{fcpl1}
\\
&& F^i_{\rm ne} = \cN_\rn  f^i_{\rm Dn} = \rho_\rn \cW_\rn \cR_\rn (v^i_\re -u^i_\rn)
 \label{fcpl2}
 \\
&& F^i_{\rm pe} = \cN_\rp  f^i_{\rm Dp}  = \rho_\rp \cW_\rp \cR_\rp (v^i_\re -u^i_\rp)
\label{fcpl3}
\eear
In these expression we can eliminate the vortex velocities $u^i_\rx$ using (\ref{velf}) and (\ref{velv}).
The final results are
\bear
&& F^i_{\rm mf} = \rho_\rn \cW_\rn \Bigg [ -\frac{\cR}{1+\cR^2} \Bigg \{ \cR f^i +
\epsilon^{ijk} \hat{\kappa}^\rn_j f_{k} + \frac{1}{\cR} \hat{\kappa}_j^\rn D^j  \hat{\kappa}^i_\rn   \Bigg \}
+ \frac{\cC_\rv}{1+\cR^2_\rp} \Bigg \{ \cR_\rp f^i_\star + \epsilon^{ijk} \hat{\kappa}_j^\rp f_{\star k}  \Bigg \}    \Bigg ]
\label{fcpl4}
\nonumber \\
\\
&& F^i_{\rm ne} = -\rho_\rn \cW_\rn \frac{\cR_\rn}{1+\cR^2} \Bigg (  \cR f^i +
\epsilon^{ijk} \hat{\kappa}^\rn_j f_{k} + \frac{1}{\cR} \hat{\kappa}_j^\rn D^j  \hat{\kappa}^i_\rn   \Bigg )
\label{fcpl5}
\\
\nonumber \\
&& F^i_{\rm pe} = - \rho_\rp \cW_\rp \frac{\cR_\rp}{1+ \cR_\rp^2}
\Bigg (   \cR_\rp f^i_\star + \epsilon^{ijk} \hat{\kappa}_j^\rp f_{\star k} \Bigg )
\label{fcpl6}
\eear


\section{Summary: The complete set of MHD equations}
\label{sec:final}

At this point we have completed our analysis of the problem of a mixed neutron superfluid and a
type~II superconductor. We have discussed the issues that arise due to the presence of vortices and magnetic fluxtubes, and accounted for the 
resulting forces in the macroscopic
equations of motion. To conclude the discussion, we will now collate the information into a ``complete'' set of MHD equations.

The final results from the previous section are rather complex. However, the model retains force terms that will be very small 
under most circumstances. In many realistic situations the forces can be simplified considerably leading to a more 
practical MHD ``toolkit''. This simplified model is presented below while, for completeness, the general equations 
are given in Appendix~\ref{app:fullMHD}.

In general, we have arrived at the following set of equations; First of all,
provided that the constituents are individually conserved (i.e. ignoring reactions), three
continuity equations relate the number densities $n_\rx$ to the transport velocities $v_\rx^i$;
\be
\partial_t n_\rx + \nabla_i ( n_\rx v^i_\rx) = 0
\ee
for neutrons (x=n), protons (p) and electrons (e). We also have
the two Euler equations \eqref{eulern}
and \eqref{eulerp}, that govern the evolution of the ``momenta''.
These Euler equations take the form
\bear
&&  (\partial_t + v^j_\rn \nabla_j )[ v^i_\rn + \en w^i_{\rm pn}] + \nabla^i \left ( \tilde{\mu}_\rn  + \Phi \right )
+ \en w_{\rm pn}^j \nabla^i v_j^\rn = f^i_{\rm mf} + \frac{1}{\rho_\rn} t^i_\rn -\nabla^i \tilde{\zeta}_\rn
\label{eulern_final}
\\
\nonumber \\
&& (\partial_t + v^j_\rp \nabla_j ) [ v^i_\rp - \ep w^i_{\rm pn} ] + \nabla^i \left ( \tilde{\mu} + \Phi \right )
- \ep w_{\rm pn}^j \nabla^i v_j^\rp   = - \frac{\rho_\rn }{\rho_\rp} f^i_{\rm mf} + f^i_{\rm mag}
\label{eulerp_final}
\eear
where $w_\mathrm{xy}^i=v_\rx^i-v_\ry^i$.

Let us now consider possible simplifying assumptions.
First of all, we know that the London field vanishes identically if the electrons are 
constrained to move with the protons, i.e. when $w_{\rp\re}^i=0$. In general,
we see from the estimate (\ref{bL_num})  that $b_\mathrm{L} \ll H_{c1},H_{\rm vn}$ and hence
we can neglect the London field contribution to the forces $t^i_\rx$ in the momentum
equations in most cases. A similar argument shows that we can use
\be
\cW_\rp^i \approx a_\rp B^i \quad \to \quad B^i \approx \cN_\rp \phi_0 \hat{\cW}^i_\rp 
\ee
where $a_\rp = e/mc$ and $\phi_0=hc/2e$. Recall that we have defined
\bear
&&\cW^i_\rn =  \epsilon^{ijk} \nabla_j  [ v^i_\rn + \en w^i_{\rm pn}] = \cN_\rn \hat{\cW}_\rn^i
\\
\nonumber \\
&& \cW^i_\rp = \epsilon^{ijk} \nabla_j [ v^i_\rp - \ep w^i_{\rm pn}] +  a_\rp B^i = \cN_\rp \hat{\cW}_\rp^i
\eear
where $\cN_\rn$ and $\cN_\rp$ are the neutron and proton vortex densities (per unit area).
With these approximations we obtain a significantly simplified set of MHD equations.
First, we have
\be
 t^i_\rn \approx \frac{\cW_\rn}{4\pi a_\rp} \Bigg [  \hat{\cW}^j_\rn \nabla_j \left ( H_{\rm vn} \hat{\cW}^i_\rn \right ) 
- \nabla^i H_{\rm vn} \Bigg ]  
\ee
and
\be
 t_\rp^i \approx \frac{B}{4\pi} \Bigg [ \hat{B}^j \nabla_j \left (  H_{\rm c1} \hat{B}^i \right ) -\nabla^i H_{\rm c1}   
\Bigg ]
\ee
Meanwhile, the magnetic forces are
\be
f^i_{\rm mag} \equiv {1 \over \rho_\rp} F^i_{\rm mag} 
\approx \frac{1}{\rho_\rp} t^i_\rp -\nabla^i \tilde{\zeta}_\rp \approx \frac{1}{4\pi \rho_\rp}
\Bigg [ B^j \nabla_j \left ( H_{\rm c1} \hat{B}^i \right ) -B \nabla^i H_{\rm c1} \Bigg ]
-\frac{1}{4\pi} \nabla^i \left ( B \frac{\partial H_{\rm c1}}{\partial \rho_\rp} \right )
\ee
and
\be
\frac{1}{\rho_\rn} t^i_\rn -\nabla^i \tilde{\zeta}_\rn \approx
 \frac{1}{4\pi \rho_\rn}
\Bigg [ \cW^j_\rn \nabla_j \left ( H_{\rm vn} \hat{\cW}^i_\rn \right ) - \cW_\rn \nabla^i H_{\rm vn} \Bigg ]
-\frac{1}{4\pi} \nabla^i \left ( B \frac{\partial H_{\rm c1}}{\partial \rho_\rn} \right ) 
\ee
The mutual friction follows from
\begin{multline}
 f^i_{\rm mf} \equiv {1 \over \rho_\rn} F^i_{\rm mf}  =
\cW_\rn \Bigg [ -\frac{\cR}{1+\cR^2} \Bigg \{ \cR f^i +
\epsilon^{ijk} \hat{\cW}^\rn_j f_{k} + \frac{1}{\cR} \hat{\cW}_j^\rn f^j  \hat{\cW}^i_\rn   \Bigg \}
+ \frac{\cC_\rv}{1+\cR^2_\rp} \Bigg \{ \cR_\rp f^i_\star + \epsilon^{ijk} \hat{\cW}_j^\rp f_{\star k}  \Bigg \}    \Bigg ]
\label{Fmf}
\end{multline}
with
\bear
&& f^i_\star \approx \frac{1}{\rho_\rp \cW_\rp} t^i_\rp \approx
\frac{1}{4\pi a_\rp \rho_\rp} \Bigg [ \hat{B}^j \nabla_j \left ( H_{\rm c1}  \hat{B}^i \right ) 
-\nabla^i H_{\rm c1} \Bigg ]
\\
\nonumber \\
&& f^i \approx \epsilon^{ijk} \hat{\cW}_j^\rn w_k^\rpn  + \cC \cR_\rp f^i_\star + \cC \epsilon^{ijk} \hat{B}_j f_{\star k}
+ \frac{1}{4\pi a_\rp \rho_\rn} \Bigg [ \hat{\cW}^j_\rn \nabla_j \left ( H_{\rm vn} \hat{\cW}^i_\rn \right ) 
-\nabla^i H_{\rm vn} \Bigg ]
\eear
Due to its relation to the current (Amp\'ere's law), we can  use the London field to
infer the electron flow;
\be
v_\re^i = v_\rp^i - { 1 \over 4 \pi  a_\rp \rho_\rp}  \epsilon^{ijk} \nabla_j b^\mathrm{L}_k
\ee
The London field is associated with the macroscopic rotation of the two fluids, in such a way that
\be
b_\mathrm{L}^i = - { 1 \over a_\rp} \left[ \epsilon^{ijk} \nabla_j ( v_k^\rp - \ep w_k^{\rm pn} ) 
- { \ep \over 1 - \en} \epsilon^{ijk} \nabla_j ( v_k^\rn + \en w_k^{\rm pn}) \right]
\ee 
These relations allows us to work out the equations relating the fluid and electromagnetic degrees of freedom.
They are, first of all, the induction
equation
\be
\partial_t B^i = \epsilon^{ijk} \epsilon_{klm} \nabla_j ( v^l_\re B^m )
 - \frac{1}{ a_\rp } \epsilon^{ijk} \nabla_j  f_k^\re 
\label{ind_final}
\ee
and secondly
\be
E^i = -\frac{1}{c} \epsilon^{ijk} v_j^\re  B_k
-\frac{1}{e} \nabla^i \left ( \mu_\re + m_\re \Phi \right ) + \frac{1}{c a_\rp} f^i_\re
\ee
In these equations $f^i_\re $ represents the total mutual friction force between the electrons the other fluids.
This can be approximated as
\be
f^i_\re \equiv -\frac{1}{\rho_\rp} \left ( F^i_{\rm pe} + F^i_{\rm ne} \right )   
\approx \frac{1}{4 \pi \rho_\rp} \frac{\cR_\rp}{1+\cR_\rp^2} \Bigg [ \cR_\rp \left \{  B^j \nabla_j (H_{\rm c1} \hat{B}^i ) 
-B \nabla^i H_{\rm c1} \right \} 
- \epsilon^{ijk} B_j \nabla_k H_{\rm c1} +  H_{\rm c1} \epsilon^{ijk} \hat{B}_j B^l \nabla_l \hat{B}_k \Bigg ]
\ee
since, in a typical neutron star, $\cN_\rp \gg \cN_\rn$.

The system of equations is closed by the Poisson equation
$\nabla^2 \Phi = 4\pi G \rho \approx 4\pi G(\rho_\rp+\rho_\rn)$
for the gravitational potential $\Phi$, the remaining Maxwell equation $\nabla_i  B^i = 0$
and by supplying the energy functional $ \cE $ (the equation of state) that is needed to
determine the chemical potentials $\tilde{\mu}_\rx$ and the entrainment parameters $\varepsilon_\rx$.

These relations complete the simplified model, which should be adequate for most problems of astrophysical 
interest.


\section{Concluding discussion}
\label{sec:conclusions}

In this paper we have developed the equations that govern the magnetohydrodynamics of superfluid and superconducting
neutron stars. To be precise, our formalism is relevant for the conditions expected in the outer core of
young and mature neutron stars where the matter composition is dominated by neutrons, protons and electrons.
The formalism does not apply in the deep core, where the matter composition could be significantly different,
with the likely presence of various hyperon species or deconfined quarks. Neither is our analysis relevant in regions of the star where
proton superconductivity is not of type II, see the discussion in Section~\ref{sec:supercon}.

In many ways, our analysis represents an upgrade of previous work on this subject, e.g. \citet{mendell98}.
First of all, we have addressed a conceptual issue regarding the dependence of the intrinsic neutron and proton vortex
 energies on the London field. Secondly, we have shown  how the tension force due to
the proton vortices (the fluxtubes) becomes the dominant magnetic force in the equations of motion while the standard Lorentz force
is eliminated. We have also provided detailed derivations of the non-dissipative forces acting on the neutron and
proton vortex arrays.
Finally, we have accounted for the mutual friction interaction between the vortex arrays
and the fluids by introducing phenomenological drag forces (in the usual way).

The multifluid formalism developed in this paper is intended to provide the ``next generation'' of neutron star magnetohydrodynamics. 
It can be applied to a number of interesting problems related to neutron star dynamics. Obvious 
possibilities involve the determination of equilibrium configurations of superconducting stars and the formulation of perturbation equations to study the oscillations of such systems. We have already
discussed the properties of magnetar oscillations, and the potentially important effect
of superfluidity/superconductivity on the quasi-periodic oscillation frequencies observed during the so-called giant
flares, elsewhere \citep{shortmagnetar}. There are many other exciting problems to consider.
Multifluid systems are known to be prone to dynamical ``two-stream''  instabilities provided there is
a sufficient degree of relative motion between the fluid components (see, for example, \citet{2stream,lars}). Recent work
for non-magnetic superfluid neutron star models has demonstrated the potential importance of such
instabilities for neutron star glitches and precession \citep{prec1,glitch}. Two-stream instabilities should also be
present in the more realistic superfluid/superconducting neutron star model discussed in this paper.
The nature of such instabilities need to be considered in detail. Another important application of our
formalism would concern the secular evolution of the magnetic field in the outer core of a neutron star.
Even a basic understanding of the magnetic field evolution in magnetar cores
could tell us a lot about the way magnetic energy is distributed in the interior as the star ages and perhaps unveil
the mechanism responsible for triggering the giant flares observed in these objects.

Further work should also expand our formalism in two important directions. We need to consider;
(i)  type I superconductivity and the transition regions between type II and type I/normal matter domains, and
(ii) the magnetohydrodynamics of the inner core in the presence of exotic matter. These are all very challenging problems
 which we hope to address in the future.


\section*{Acknowledgments}

NA acknowledges support from STFC via grant number PP/E001025/1. KG is supported by the German Science
Foundation (DFG) via SFB/TR7. LS is supported by the European Research Council under Contract No.\ 204059-QPQV, 
and the Swedish Research Council under Contract No.\ 2007-4422


\appendix

\section{Individual neutron and proton vortices}
\label{app:carter}

In this Appendix we work out the fluid flows and magnetic fields associated with individual neutron and proton vortices, 
accounting for the entrainment in the appropriate way. We then derive the expression for the energy
(per unit length) associated with each object. The results form a key part of the derivation of the macroscopic
forces in Section~\ref{sec:variational}.

Some of these results are well known; they can be found in, for example, \citet{mendell} and \citet{ruderman}.
However, some aspects of our discussion have not been sufficiently exposed in the literature. 
In particular, we show that the energy of a single vortex/fluxtube is independent of the macroscopic London field. 
This result was first pointed out by \citet{carter} in the context of a simple model of a uniformly rotating superconductor with a 
single fluxtube placed along the rotation axis. A more general derivation for a mixture of superfluids was subsequently given 
by \citet{prix00}. Our analysis, which is focused on the case of a system composed of superfluid neutrons, superconducting protons 
and electrons, provides a simpler and more intuitive derivation of this result.


\subsection{The flow and magnetic field for an isolated neutron/proton vortex}

\label{app:fstraight}

Let us consider the case of an single straight neutron or proton vortex.
At the mesoscopic level (indicated by bars on the relevant quantities), we have
\be
{ 1 \over m } \epsilon^{ijk} \nabla_j \bar{p}^\rn_k  = \kappa^i_\rn \delta (\vec{r}), \qquad
{ 1 \over m } \epsilon^{ijk} \nabla_j \bar{p}^\rp_k + a_\rp \bar{B}^i_\rx = \kappa^i_\rp \delta (\vec{r})
\label{eq1}
\ee
where $\kappa_\rp =0$ for a neutron vortex and $\kappa_\rn=0$ for a proton vortex. 
The argument of the two-dimensional delta function can be expressed in terms of the (cylindrical) radial
distance $r$ away from the vortex core. In our derivation we will not consider a detailed model for the vortex core,
but focus our attention on the behaviour at distances larger than (say) the coherence length $\xi_\rx$.
In essence, we consider a simple hollow-core model.

Starting from (\ref{eq1}) we can immediately solve for the canonical momenta
\be
\bar{\pi}^i_\rn = \bar{p}_\rn^i = { m \kappa_\rn \over 2 \pi r} \hat{\varphi}^i,
\qquad
\bar{\pi}^i_\rp = \bar{p}_\rp^i + \frac{e}{c} \bar{A}^i_\rx =  { m \kappa_\rp \over 2 \pi r} \hat{\varphi}^i
\label{meso_pi}
\ee
where the magnetic potential is defined in the usual way, i.e. $\bar{B}^i_\rx = \epsilon^{ijk} \nabla_j \bar{A}_k^\rx$.

In order to obtain the proton fluid momentum we first need to calculate the local magnetic field $\bar{B}^i_\rx$.
To this end we need to consider Amp\'ere's law
\be
\epsilon^{ijk} \nabla_j \bar{B}^\rx_k = { 4 \pi \over c} e n_\rp \bar{v}_\rp^i
\label{meso_ampere}
\ee
where the total current is supplied by the proton flow only. Taking the curl of (\ref{meso_ampere}) and
using $\nabla_j \bar{B}_\rx^j=0 $ we obtain
\be
\nabla^2 \bar{B}_\rx^i - { 1 \over \Lambda^2_*} \bar{B}^i_\rx  =
- {\phi_\rx \over \Lambda^2_*} \hat{\kappa}_\rx^i \delta (\vec{r})
\label{mesoB}
\ee
where we have defined the flux parameters
\be
\phi_\rp = \phi_0 = \frac{hc}{2e}, \qquad \phi_\rn = -\frac{\ep}{1-\en} \phi_0
\ee
and the London penetration length
\be
\Lambda^2_*= \frac{\epstar}{4\pi\rho_\rp} \left ( \frac{mc}{e} \right )^2
\ee
The entrainment parameter $\epstar$ is defined in eqn. (\ref{gam}). 
This expression leads to the  estimate (\ref{Lambda}) used in Section~\ref{sec:supercon}.

The solution of (\ref{mesoB}) is
\be
\bar{B}_\rx^i =  {\phi_\rx \over 2 \pi  \Lambda^2_*} K_0 \left( { r \over \Lambda_*} \right) \hat{\kappa}_\rx^i
\label{mesoBsol}
\ee
where $K_0$ is an associated Bessel function. It is straightforward to calculate the magnetic flux
through a disk surface of radius $r_{\rm max} \gg \Lambda_*$,  perpendicular with respect to $\kappa^i_\rx$.
The relevant surface integral is
\be
\int_{\rm disk}  \bar{B}^i_\rx dS  = \phi_\rx \hat{\kappa}_\rx^i  
\ee
For a proton vortex this result represents the expected flux quantisation.

Going back to (\ref{meso_ampere}) we can obtain the proton velocity (for both vortex types)
\be
\bar{v}^i_\rp = \frac{\kappa}{2\pi \epstar \Lambda_*} \frac{\phi_\rx}{\phi_0} K_1 \left( { r \over \Lambda_*} \right)
\hat{\varphi}^i
\ee
In other words $\bar{v}_\rp^i \sim \exp(-r/\Lambda_*)$ sufficiently far from the vortex axis.
From the definition
\be
\bar{p}^i_\rx = m [ (1 -\varepsilon_\rx) \bar{v}^i_\rx + \varepsilon_\rx \bar{v}_\ry^i]
\ee
we find that
\be
\bar{v}_\rp^i = { 1 \over m \epstar }  \left[ \bar{p}_\rp^i -\frac{\ep}{1-\en} \bar{p}_\rn^i \right]
\label{velp}
\ee
Then it follows that at a distance $r \gg \Lambda_*$,
\be
\bar{p}_\rp^i \approx \frac{\ep}{1-\en} \bar{p}_\rn^i
\label{pcond}
\ee
for both neutron and proton vortices.

We also need to calculate the potential $\bar{A}^i_\rx$. The symmetry of the problem dictates that
$\bar{A}^i_\rx = \bar{A}_\rx (r) \hat{\varphi}^i$ and we end up with a simple differential equation
\be
\frac{1}{r} \partial_r ( r \bar{A}_\rx ) = \frac{\phi_\rx}{2\pi\Lambda^2_\star}  K_0 \left( { r \over \Lambda_*} \right)
\ee
the solution of which is
\be
\bar{A}^i_\rx = \frac{\phi_\rx}{2\pi} \left [ \frac{c_1}{r} -\frac{1}{\Lambda_\star} K_1 \left( { r \over \Lambda_*} \right)  \right ]
\hat{\varphi}^i
\ee
where $c_1$ is a constant.

For a proton vortex we obviously have $\bar{p}^i_\rn =0$. Combining this with the previous results for $\bar{\pi}^i_\rp$
and $\bar{A}^i_\rp$ we find that
\be
\bar{p}^i_\rp = \frac{m\kappa}{2\pi} \left [ \frac{1-c_1}{r} +
\frac{1}{\Lambda_\star} K_1 \left( { r \over \Lambda_*} \right) \right ] \hat{\varphi}^i
\to \frac{m\kappa}{2\pi \Lambda_\star} K_1 \left( { r \over \Lambda_*} \right)  \hat{\varphi}^i
\qquad (\mbox{proton vortex})
\ee
after fixing $c_1 =1$, in order to ensure consistency with the condition (\ref{pcond}). Note that, we could
have arrived at the same result without involving the vector potential by simply using eqn.~(\ref{velp}), which for a proton vortex
reduces to $\bar{p}^i_\rp= m\epstar \bar{v}^i_\rp $.

For a neutron vortex we have $\bar{\pi}^i_\rp =0$ and therefore
\be
\bar{p}^i_\rn = \frac{m \kappa}{2\pi r} \hat{\varphi}^i, \qquad
\bar{p}^i_\rp = -\frac{e}{c} \bar{A}^i_\rn =
\frac{m\kappa }{2\pi} \frac{\ep}{1-\en} \left [ \frac{1}{r} -\frac{1}{\Lambda_\star} K_1 \left( { r \over \Lambda_*} \right)  \right ]
\hat{\varphi}^i
\quad (\mbox{neutron vortex})
\ee
where (as before) we have fixed $c_1 =1$ in order to ensure that (\ref{pcond}) is satisfied.


\subsection{The vortex energy}
\label{app:Evort}

Let us now discuss the energy ${\cal E}_{\rm vx} $ of a single vortex (either associated with the neutrons or the protons) immersed in given
macroscopic fluid flows $v_\rx^i $. The analysis is carried out in the vortex rest-frame.
In order to work out the energy associated with the vortex we need the difference between the localized
magnetic/fluid energy and the total energies.
Thus we write the  {\it total} local momentum  as the sum of $\bar{p}^i_{\rx} $ and the (smooth) macroscopic momentum $p^i_\rx $.
We then have $ p^i_\rx = m_\rx \left ( {v}^i_\rx -u^i_\rv + \varepsilon_\rx {w}^i_\ryx \right )$, where $u^i_\rv $ is the
vortex velocity in the inertial frame. The total local magnetic field can be written in a similar way, as the sum
of $\bar{B}^i_\rx $ plus the (locally smooth) London field,
\be
B^i_{\rm loc} =\bar{B}^i_\rx + b^i_{\rm L}
\ee
Since $ \bar{B}_\rx^i $ behave as $\sim e^{-r/\Lambda_*} $ sufficiently away from the vortex axis (see eqn.~(\ref{mesoBsol}))
we have the asymptotic behaviour $B_{\rm loc}^i \to b^i_{\rm L} $.

The total vortex energy $\cE_{\rm vx} $ per unit length is the sum of its magnetic and kinetic
energies plus the ``condensation energy'' associated with the normal matter in the vortex core:
\be
\cE_{\rm vx} = \cE_{\rm mag} + \cE_{\rm kin} + \cE_{\rm con}
\ee
For the present discussion it will suffice to focus on the first two terms
which we write as integrals over the magnetic and kinetic energy densities,
\be
\cE_{\rm mag} = \int  {E}_{\rm mag} dS
\label{Emag1}
\ee
\be
\cE_{\rm kin} = \int  {E}_{\rm kin} dS
\label{Ekin1}
\ee
The integrals in these expressions are taken over the surface of a disk perpendicular to the vortex axis and
extending to a radius $r \sim d_\rx $, where $d_\rx$ is the distance to the nearest vortex.

For the magnetic energy we get
\be
{E}_{\rm mag} = \frac{B^2_{\rm loc}}{8\pi} - \frac{b^2_{\rm L}}{8\pi}  = \frac{\bar{B}^2_\rx}{8\pi}
+ \frac{1}{4\pi} b^i_{\rm L} \bar{B}_i^\rx
\label{Umag1}
\ee
That is,
\be
\cE_{\rm mag} = \int \frac{\bar{B}^2_\rx}{8\pi} dS  + \frac{1}{4\pi} b^i_{\rm L}
\int  \bar{B}_i^\rx dS = \int \frac{\bar{B}^2_\rx}{8\pi}  dS + \frac{\phi_\rx}{4\pi} (\hat{\kappa}_i^\rx
b^i_{\rm L})
\label{Emag2}
\ee

The kinetic energy is calculated following the same prescription. Before we can do this, we
need to clarify how the kinetic energy is defined for a multifluid system with constituents coupled through
entrainment.  Within the canonical formalism of \citet{prix04} it can been shown that the Hamiltonian
density associated with the hydrodynamical Lagrangian $\cL_{\rm H} $ is given by the standard Legendre
transformation ${\cal H}_{\rm H} = \sum_\rx n^i_\rx p_i^\rx -\cL_{\rm H} $. Hence, we  define the
 kinetic energy density as
\be
{E}_{\rm kin} = \frac{1}{2} ( \cL_{\rm H} + {\cal H}_{\rm H} ) = \frac{1}{2} \sum_\rx n^i_\rx p_i^\rx
\ee
On the macroscopic scale, and in terms of the individual momenta, this leads to
\be
E_\mathrm{kin} = { 1 \over 2 m^2} (1 - \en - \ep)^{-1} \left[\, \rho_\rn (1-\ep) p_\rn^2
- (\rho_\rn \en + \rho_\rp \ep) p_\rp^i p^\rn_i + \rho_\rp (1- \en) p_\rp^2 \, \right]
\ee
In order to single out the vortex contribution we take the difference between this expression evaluated for the total
momenta, $p_\rx^i + \bar{p}_\rx^i$, and the expression we get using only the momenta associated with the vortex itself,
$\bar{p}_\rx^i$. After some manipulations, where it is useful to express the London field
in terms of a vector potential $b^i_{\rm L} = \epsilon^{ijk} \nabla_j A^{\rm L}_k $, i.e. use
\be
A^i_{\rm L} =  -\frac{1}{m a_\rp} \left (\, p^i_\rp -\frac{\ep}{1-\en} p^i_\rn
\, \right )
\label{alphapot}
\ee
it follows that
\begin{multline}
E_{\rm kin} =  \frac{1}{2m^2} (1-\en-\ep)^{-1} \left [ \, \rho_\rn (1-\ep) \bar{p}^2_{\rn}
+ \rho_\rp  (1-\en) \bar{p}^2_{\rp} - 2\rho_\rn \en \bar{p}^i_{\rn} \bar{p}_i^{ \rp} \,
\right ] 
\\
-\frac{1}{4\pi} b^i_{\rm L} \bar{B}_i^\rx  + \frac{\rho_\rn}{m^2 (1-\en)} p_i^\rn \bar{p}^i_{ \rn}
-\frac{1}{4\pi} \epsilon^{ijk} \nabla_j ( \bar{B}_k^\rx A_i^{\rm L} )
\end{multline}
When we work out the  surface integral of ${E}_{\rm kin} $ we find that the last two terms
vanish due to the symmetry of the problem. Hence, the final expression for the kinetic energy (per unit vortex length)
is,
\be
\cE_{\rm kin} = \frac{1}{2m^2} (1-\en-\ep)^{-1} \left [ \rho_\rn (1-\ep) \int  \bar{p}^2_{\rn} dS
+ \rho_\rp (1-\en) \int  \bar{p}^2_{ \rp} dS - 2 \rho_\rn \en  \int \bar{p}^i_{\rn} \bar{p}_i^{\rp}  dS
\right ]
- \frac{\phi_\rx}{4\pi} (\hat{\kappa}_i^\rx b^i_{\rm L})
\label{Ekin2}
\ee
Each of the partial energies (\ref{Emag2}) and (\ref{Ekin2}) show an explicit dependence on the London field.
However, these London field terms cancel exactly in the sum $\cE_{\rm kin} + \cE_{\rm mag} $.
The final result for the total vortex energy is therefore
\be
\cE_{\rm vx} = \frac{1}{2m^2} (1-\en-\ep)^{-1} \left [ \rho_\rn (1-\ep) \int  \bar{p}^2_{\rn} dS
+ \rho_\rp (1-\en) \int  \bar{p}^2_{ \rp} dS - 2 \rho_\rn \en  \int \bar{p}^i_{\rn} \bar{p}_i^{\rp}  dS
\right ]
+ \frac{1}{8\pi} \int  \bar{B}^2_\rx dS  + \cE_{\rm con}
\label{Energ_vort}
\ee
which is clearly independent of the London field and the macroscopic velocities. This is in agreement
with the conclusions of \citet{carter} and \citet{prix00}.

Inserting the   results for the vortex flow and the magnetic field in the energy formula 
(and ignoring the condensation energy) we arrive at
\be
\cE_{\rm vp} = \frac{\kappa^2 \rho_\rp}{4\pi \varepsilon_\star} \log \left ( \frac{\Lambda_\star}{\xi_\rp}\right )
\label{evortp}
\ee
and
\be
\cE_{\rm vn} \approx {\kappa^2 \over 4 \pi} { \rho_\rn  \over 1-\en}  \log \left( \frac{d_\rn}{\xi_\rn} \right)
\label{evortn}
\ee
In both cases the vortex energy is dominated by the kinetic term (the first term in (\ref{Energ_vort})).

Finally, we  also obtain the critical field (defined by eqn.~(\ref{Hvx_def}))
\be
H_{\rm c1} = \frac{4\pi \cE_{\rm vp}}{\phi_0} \quad \to \quad 
H_{\rm c1} = \frac{\kappa \rho_\rp}{\varepsilon_\star} \log \left ( \frac{\Lambda_\star}{\xi_\rp} \right )
\ee
This expression leads to the numerical estimate (\ref{Hc1def}) of Section~\ref{sec:supercon}.
Ignoring the slow logarithmic dependence, this can be written as
\be
H_{\rm c1} (\rho_\rp,\rho_\rn) = h_{\rm c} \frac{\rho_\rp}{\epstar}, \qquad h_{\rm c} \approx \mbox{constant}
\ee
which is a form suitable for use in the final MHD equations of Section~\ref{sec:final}. 


\section{The full MHD equations}
\label{app:fullMHD}

In this Appendix we present our final equations for the full MHD model, without the
simplifications described in Section~\ref{sec:final}.

To begin with, and provided that constituents are individually conserved, we have the 
three continuity equations
\be
\partial_t n_\rx + \nabla_i ( n_\rx v^i_\rx) = 0
\ee
for neutrons (x=n), protons (p) and electrons (e). We also have two equations that govern the evolution of the momenta. 
These Euler equations take the form
\bear
&&  (\partial_t + v^j_\rn \nabla_j )[ v^i_\rn + \en w^i_{\rm pn}] + \nabla^i \left ( \tilde{\mu}_\rn  + \Phi \right )
+ \en w_{\rm pn}^j \nabla^i v_j^\rn = f^i_{\rm mf} + \frac{1}{\rho_\rn} t^i_\rn -\nabla^i \tilde{\zeta}_\rn
\\
\nonumber \\
&& (\partial_t + v^j_\rp \nabla_j ) [ v^i_\rp - \ep w^i_{\rm pn} ] + \nabla^i \left ( \tilde{\mu} + \Phi \right )
- \ep w_{\rm pn}^j \nabla^i v_j^\rp   = - \frac{\rho_\rn }{\rho_\rp} f^i_{\rm mf} + f^i_{\rm mag}
\eear
where $w_\mathrm{xy}^i=v_\rx^i-v_\ry^i$.

The various force terms are given by
\be
 t^i_\rn = \frac{1}{4\pi a_\rp} \cW^j_\rn \Bigg \{ \,   \nabla_j  ( H_{\rm vn} \hat{\cW}_\rn^i) -  \nabla^i  ( H_{\rm vn} \hat{\cW}^\rn_j)
+  \nabla_j \left( \frac{\ep}{1-\en} b^i_{\rm L} \right)
- \nabla^i \left(  \frac{\ep}{1-\en} b_j^{\rm L} \right )  \, \Bigg \}
\ee
and
\be
f^i_{\rm mag}  \equiv {1 \over \rho_\rp} F^i_{\rm mag} 
=\frac{1}{4\pi a_\rp \rho_\rp} \Bigg\{   \cW^j_\rp \left[  \nabla_j ( H_{c1} \hat{\cW}_\rp^i) - \nabla^i (H_{c1} \hat{\cW}^\rp_j) \right]
-\left ( \nabla_j b^i_{\rm L} - \nabla^i b_j^{\rm L} \right )  \epsilon^{jkl} \nabla_k  [ v_l^\rp - \ep w_l^{\rm pn} ]  \Bigg\}
-\nabla^i \tilde{\zeta}_\rp
\ee
where $a_\rp = e/mc$ and
\be
\tilde{\zeta}_\rx = { 1 \over 4 \pi a_\rp} \left[  \frac{\partial H_{\rm vn}}{\partial \rho_\rx} \cW_\rn
+ \frac{\partial H_{c1}}{\partial \rho_\rx} \cW_\rp + {\partial \over \partial \rho_\rx} \left( \frac{\ep}{1-\en} \right)
\cW^\rn_i b_\mathrm{L}^i  \right]
\ee
The canonical vorticities $\cW^i_\rx$ were defined as
\bear
&&\cW^i_\rn =  \epsilon^{ijk} \nabla_j  [ v^i_\rn + \en w^i_{\rm pn}] = \cN_\rn \hat{\cW}_\rn^i
\\
\nonumber \\
&& \cW^i_\rp = \epsilon^{ijk} \nabla_j [ v^i_\rp - \ep w^i_{\rm pn}] +  a_\rp B^i = \cN_\rp \hat{\cW}_\rp^i
\eear
where $\cN_\rn$ and $\cN_\rp$ are the neutron and proton vortex densities (per unit area).

Recall also that the total magnetic field $B^i$ and the London field $b^i_{\rm L}$ are given by
\bear
&&B^i  = b_{\rm L}^i +  \phi_0 \left( \cW_\rp^i  -   \frac{\ep}{1-\en} \cW_\rn^i \right)
\\
\nonumber \\
&& b_\mathrm{L}^i = - { 1 \over a_\rp} \left[ \epsilon^{ijk} \nabla_j ( v_k^\rp - \ep w_k^{\rm pn} ) 
- { \ep \over 1 - \en} \epsilon^{ijk} \nabla_j ( v_k^\rn + \en w_k^{\rm pn}) \right]
\eear
where $\phi_0 = hc/2e$. 

The mutual friction follows from
\be
 f^i_{\rm mf} \equiv {1 \over \rho_\rn} F^i_{\rm mf}  =
\cW_\rn \Bigg [ -\frac{\cR}{1+\cR^2} \Bigg \{ \cR f^i +
\epsilon^{ijk} \hat{\cW}^\rn_j f_{k} + \frac{1}{\cR} \hat{\cW}_j^\rn f^j  \hat{\cW}^i_\rn   \Bigg \}
+ \frac{\cC_\rv}{1+\cR^2_\rp} \Bigg \{ \cR_\rp f^i_\star + \epsilon^{ijk} \hat{\cW}_j^\rp f_{\star k}  \Bigg \}    \Bigg ]
\ee
with
\bear
&& f^i = \epsilon^{ijk} \hat{\cW}_j^\rn w^{\rm en}_k - \cC \epsilon^{ijk} \hat{\cW}^\rp_j \left  \{ \cR_\rp  w^\rpe_k +
  \epsilon_{klm} \hat{\cW}^l_\rp w^m_\rpe - \frac{1}{\rho_\rp \cW_\rp} t_k^\rp \right \}
+ \frac{1}{\rho_\rn \cW_\rn} t^i_\rn + \frac{\cC \cR_\rp}{\rho_\rp \cW_\rp} t^i_\rp
\nonumber \\
\\
&& f^i_\star =  -\epsilon^{ijk} \hat{\cW}_j^\rp w^{\rm pe}_k + \frac{1}{\rho_\rp \cW_\rp } t^i_\rp
\\
\nonumber \\
&& t^\rp_i = \frac{1}{4\pi a_\rp} \cW^j_\rp \left [\,    \nabla_j ( H_{c1}\hat{\cW}^\rp_i) -  \nabla_i ( H_{c1}\hat{\cW}^\rp_j)
-  \left ( \nabla_j b_i^{\rm L}
- \nabla_i b_j^{\rm L} \right ) \, \right ]
\eear
Using Amp\'ere's law, we can  use the London field to infer the electron flow;
\be
v_\re^i = v_\rp^i - { 1 \over 4 \pi  a_\rp \rho_\rp}  \epsilon^{ijk} \nabla_j b^\mathrm{L}_k
\ee
There are two more equations relating the fluid and electromagnetic degrees of freedom.
The induction equation
\be
\partial_t B^i = \epsilon^{ijk} \epsilon_{klm} \nabla_j ( v^l_\re B^m )
 - \frac{1}{ a_\rp } \epsilon^{ijk} \nabla_j  f_k^\re 
\label{ind_final2}
\ee
and 
\be
E^i = -\frac{1}{c} \epsilon^{ijk} v_j^\re  B_k
-\frac{1}{e} \nabla^i \left ( \mu_\re + m_\re \Phi \right ) + \frac{1}{c a_\rp} f^i_\re
\ee
where $f^i_\re$ is the total mutual friction force between the electrons and the other fluids.
It is given by
\be
f^i_\re \equiv -\frac{1}{\rho_\rp} \left ( F^i_{\rm pe} + F^i_{\rm ne} \right ) 
= \cW_\rn \frac{\rho_\rn}{\rho_\rp} \frac{\cR_\rn}{1+\cR^2} \Bigg [  \cR f^i +
\epsilon^{ijk} \hat{\cW}^\rn_j f_{k} + \frac{1}{\cR} \hat{\cW}_j^\rn f^j  \hat{\cW}^i_\rn  \Bigg ]
+ \cW_\rp \frac{\cR_\rp}{1+\cR^2_\rp}
\Bigg [ \cR_\rp f^i_\star + \epsilon^{ijk} \hat{\cW}_j^\rp f_{\star k} \Bigg ]
\ee
The system of equations is closed by the Poisson equation
$\nabla^2 \Phi = 4\pi G \rho \approx 4\pi G(\rho_\rp+\rho_\rn)$
for the gravitational potential $\Phi$, the remaining Maxwell equation $\nabla_i  B^i = 0$
and by supplying the energy functional $ \cE $ (the equation of state) that is needed to
determine the chemical potentials $\tilde{\mu}_\rx$ and the entrainment parameters $\varepsilon_\rx$.



\end{document}